\documentclass[a4paper]{article}

\usepackage{psfig}
\newcommand{\boldsymbol}[1]{\mbox{\boldmath $#1$}}
\newcommand{\dfrac}[2]{\displaystyle \frac{#1}{\displaystyle #2}}

\title{An evolutionary model for simple ecosystems}
\author{{\Large Franco Bagnoli}\footnote{Also INFM and INFN, sez. di
Firenze, Largo E. Fermi, 2 I-50125 Firenze, Italy}\\
Dipartimento di Matematica Applicata, Universit\`a di Firenze\\
Via S. Marta, 3 I-50139 Firenze, Italy\\
bagnoli@dma.unifi.it\\ 
and \\
{\Large Michele Bezzi}\\
SISSA - Programme in Neuroscience\\
Via Beirut 2-4 I-34103, Trieste, Italy\\
michele@sissa.it
}

\begin{document}
\maketitle

\begin{abstract}
In this review some simple models of asexual populations evolving on smooth
landscapes are studied. The basic model is based on a cellular automaton, 
which is analyzed here  in the  spatial mean-field limit. Firstly,
 the evolution
on a fixed fitness landscape is considered. The
correspondence between the time evolution of the population and equilibrium
properties of a statistical mechanics system is investigated, finding the limits
for which this mapping holds. The mutational meltdown, Eigen's 
error threshold and  Muller's ratchet phenomena are studied in the framework of
a simplified model. Finally, the shape of a quasi-species and the condition of
coexistence of multiple species in a static fitness landscape are analyzed. 
In the second part, these results are applied to the study of the 
coexistence of quasi-species in the
presence of competition, obtaining the conditions for a robust speciation effect
in asexual populations.  
\end{abstract}

\section{Introduction}
\label{sec:Introduction}
The object of our study is the self organization 
of an ensemble of interacting individuals: we would like 
to develop a model of a co-evolving 
ecosystem. In this paper only the very few steps of this project are
presented; we study the emerging features of asexual populations
in very simple environments.  

This paper complements (and sometimes overlaps) 
the review by Baake and Gabriel~\cite{Baake:review}. Many of the 
mathematical topics we deal with (say, error threshold, mutational
meltdown, Muller's ratchet) find there a correct biological
perspective. We also suggest Baake and Gabriel's review as a
source of commented references.

Several models exist concerning the properties of multi-species
populations (see for instance Refs.~\cite{Maynard:multispecies,Webworld}),
however we would like to start at a more
elementary level, based on individual dynamics. 

Our basic building block is a very simple schematization of an individual:
just a string of symbols (the \emph{genotype}) corresponding to a few
phenotypic traits, which are functions of the genotype.   In this way
we model haploid individuals both without sexual reproduction and without
polymorphism.  In this paper we limit ourselves to the study of spatially
homogeneous systems (spatial mean field),
which means that  we are simulating the evolution in a well
stirred container.

In a very fundamental model, each individual would feed on somebody
else, including the external source of energy. 
A similar approach is presented in the web-world model~\cite{Webworld},
which is however
based on the species concept, rather than on individual dynamics. 
In many cases, one is interested in  only a part of an
ecosystem. Very few studies (if any) take into consideration all 
interactions
among organisms, say from mammals and plants to  the microscopic world.
One is tempted to assume that an ecosystem (which should in principle
correspond to the whole earth) can be broken into subsystems each of
which is weakly coupled to the rest of the world and has different
characteristic times.
For instance, one could focus on
animals, considering the vegetables as a static substrate. Or,
alternatively, study the vegetables considering the animals as
self-averaging fluctuations. 
This simplification is valid only for a limited amount of time, which may
be long on the individual time-scale. Eventually, a rearrangement
(say, climate change, asteroid
impact, human pollution) will heavily change the 
system. 

This focusing on a
part of the whole system can be modeled by the concept of the \emph{fitness
landscape}. This landscape is customary defined as  the average number
of offsprings which reach the reproduction age, and in general it is originated
by complex interactions with many other individuals. Since the
outside world is considered fixed, also the interactions with
individuals belonging to it do not change with time.
 Thus the reproduction rate can
be divided into a static term (the static fitness landscape) and an
interaction term, which depend on the presence of individuals in the
subsystem under study. 

The internal interactions can be classified into predation 
(parasitism), competition or cooperation, depending on how the
presence of an individual affects the reproduction rate of another
one. In a fundamental description, only predation or parasitism
should be present; however, the average over the outside world can
induce the presence of competition or cooperation between two
individuals in the same subsystem. 

The evolution on a static fitness landscape (i.e. without internal
interactions among individuals) is a very interesting
subject by itself, and will be studied in this article limited to very
smooth landscapes.
This subject is directly related to the
properties of genealogical trees. Since the landscape is fixed, the
evolution of a population can be obtained as a sum over all possible
stories of individuals, in a way similar to using a path integral
approach in equilibrium statistical mechanics. This equivalence is
investigated further in Section~\ref{sec:Path}. We find that on a flat
enough landscape and for large enough mutation rates the asymptotic
distribution is indeed a Boltzmann one, in which the fitness plays the
role of the energy and mutations that of a temperature. 

One of the most striking feature of evolution is the breakdown of an
uniform distribution (both in the genotypic and in the phenotypic
space) and the formation of the clusters of individuals that are
eventually denoted as ``species''. The definition of a species is
not a trivial task. One definition (a) is based simply on the phenotypic
differentiation among the clusters, and can be easily applied also to
asexual organisms, such as bacteria. 
Another definition (b), that applies only to sexual
organism, is based on the inter-breeding possibility. Finally (c), one can
define the species on the basis of the genotypic distance of
individuals, taking into consideration their genealogical
story~\cite{Evolution}.
  
Many properties of genealogical trees on flat landscapes have been
studied by Derrida, Higgs and
Peliti~\cite{DerridaPeliti91,HiggsDerrida91,HiggsDerrida92}, 
both for asexual and
sexual reproduction.  Their conclusion  is that species (defined using
the reproductive isolation, definition (b)) can appear in
flat static landscapes provided  with sexual reproduction and
discrimination of mating. In some sense these authors have identified
definitions (b) and (c). 

Also the reason for the existence of species is quite
controversial~\cite{Maynard:Genetic}. We are here referring to the
formation of species in a spatially  homogeneous environment, i.e.\ to
\emph{sympatric} speciation. In this frame of reference, a niche is
 a phenotypic realization of relatively high fitness.
 Species have obviously to do with niches, but one cannot assume that 
 the coexistence of 
species simply reflects the presence of ``pre-existing'' 
niches; on the contrary,
what appears as a niche to a given individual is co-determined by the 
presence of  other individuals (of the same or of different species).
In other words, niches are the product of co-evolution.

Most authors consider sexual
discrimination as the fundamental ingredient for the formation of
species in flat (or almost flat) fitness landscapes (no niches).     
In this paper we deal only with asexual populations, so we stick to
the phenetic definition (a). Instead of the ``species'' term 
(which
 has a very precise meaning for biologists~\cite{Gabriel})
we adopt the weaker term
``quasi-species'', to indicate a group of phenotypically similar
individuals, separated (again, in the phenotypic space) by other
groups. We further use  the term ``speciation'' to denote the process of
division of a quasi-species in two or more groups, or,
considering a quasi-static situation, the coexistence of several
quasi-species.  

The speciation phenomenon  can occur  on static landscapes with
several (almost degenerate) maxima, which correspond to the niche
concept. However, as discussed in Section~\ref{sec:Coexistence}, in
this case  the coexistence depends sensitively on the mutation rate,
or is just a transient phenomenon. On very rough landscapes, this
transient behavior can extend to very long times, in a way similar to
the spin glass behavior~\cite{Anderson,Peliti:SpinGlass}.   

The roughness of the landscape depends also on the observation scale,
and is minimal on the individual level and much larger on the species
level. Since we are interested in the individual dynamics, we assume
quite smooth landscapes. 

One of the first studies of evolution an a static landscape
was performed by Eigen and others~\cite{Eigen71,Eigen:quasispecies}. They investigated the role of mutation in
destroying a quasi-species located around the fittest strain. This
effect is called the error threshold transition, and takes the form of
a real phase transition (first order) in the limit of infinite genome
length and infinite population~\cite{Galluccio}. The Error threshold
for finite populations has been studied in
Refs.~\cite{NowakSchuster,Baake,Alves}.
 This phase transition is not present for smooth landscapes (for an
 example of a study of evolution on a smooth landscape, see
 Ref.~\cite{Kessler}). 
The limit of long genomes is interesting, since in this case
one can neglect to consider back mutations~\cite{Baake1}. 
In this case another 
mechanism is present, the so called Muller's
ratchet~\cite{Lynch90,Lynch93}
or stochastic
escape~\cite{Higgs95,Wood96}, which, for
finite populations, causes the loosing of fitter strains by stochastic
fluctuations. Since it relies on a random process, 
the average escape time is
of the order of the exponential of the size of the
strain~\cite{Pal}, 
and thus this effect is relevant only for small
populations. 

On the other hand, this effect can lower the average fitness of the
population below a viable threshold, after which extinction can
occur~\cite{Malarz}
while the distribution still retains the quasi-species bell shape. Both
the error threshold and the Muller's ratchet phenomenon will be
investigated in Section~\ref{sec:Toy}.

In Section~\ref{sec:DynamicalEvolution}, we include the effects of the
interactions among individuals belonging to the subsystem under study.
The most fundamental terms are two-body interactions. 
In this paper we are mainly interested in the speciation phenomenon,
i.e.\ to the differentiation of otherwise similar individuals.  
Intra-specific cooperation, i.e.\ the increase of reproduction rate due
to the presence of a genetically related individual is presumably due
to familiar or group structure, and will not be considered here. 

We assume that the more similar two individuals are, the
greater is the competition term that lowers their fitness. One can think
that this competition term arises from averaging over shared resources
belonging to the outside world. 
 
One of the most interesting feature that originates from the
inclusion of an intra-specific competition term is phenotypic
differentiation.  Let us assume that the static fitness landscape has
a single smooth maximum. Without competition, the distribution of  an
asexually reproducing population would follow a bell shaped curve,
peaked around the maximum and whose width is given by mutation and
also by the curvature of the maximum (for small mutation rates).  If
the competition term is large enough, this single peak curve will
split into several ``quasi-species'', separated by the range of the 
inter-specific competition, as shown in
Section~\ref{sec:PhenotypicSpeciation}.  

The fact that one can observe phenotypic splitting of a population
solely due to competition is very interesting. In presence of sexual
reproduction (which has not been studied here) 
the competition lowers, since the phenotypic distribution is broader.
This could well be a valid reason for the evolutionary survival of sexual
reproduction in spite of its cost~\cite{costofsex,Maynard}. 
Moreover, if the splitting due to
competition holds even in presence of sexual reproduction (under study
at present), it may be invoked as an effective mechanism 
for sympatric speciation.

Mutations act on the
genotype, while selection acts on the phenotype. A single mutation can
greatly alter the phenotype. However, if one neglects immediately lethal
mutations (they simply change the fitness of the genome), most of
surviving mutants have very similar phenotypes with respect to the
parent. Thus, in many cases
one can get rid of the genotypic space: for large
scale evolution (formation of species) one can assume that mutations
induces a diffusion in the phenotypic space, while for small scale
evolution (asymptotic distribution of intra-specific traits) one can
assume that mutations are able to generate all possible phenotypic
traits.  

The outline of this paper is the following. In the next Section we
discuss the ingredients needed in our models. 
In Section~\ref{sec:CA} we present a cellular automaton model of
an evolving ecosystem which is our microscopic reference system. 
The spatial mean field theory of it allows us to obtain
(Section~\ref{sec:MF})
the well known mutation-selection equation (corresponding to a
 reaction-diffusion process in the genotypic space). 
 We then develop some
considerations about evolving ecosystems without competition: 
in Section~\ref{sec:Path} we examine to what extent 
an evolutionary system can be considered equivalent to an equilibrium
statistical model; in
Section~\ref{sec:Toy} we present a toy model exhibiting the error threshold, 
the mutational meltdown~\cite{Lynch90,Bernardes}  and  the
Muller's ratchet effects; finally, in Section~\ref{sec:PhenotypicEvolution}
we derive the shape of a quasi-species as a function of the height and
curvature of a maximum of the static fitness landscape. These last
results will be useful to determine the distribution of quasi-species
in presence of competition.
Afterward, in Section~\ref{sec:Coexistence} we analyze
to what extent coexistence can be present on a static fitness
landscape.

In Section~\ref{sec:DynamicalEvolution}, 
we introduce the intra-specific
competition interaction. 
We show that such a model can present phenotypic
differentiation (speciation: formation of quasi-species) 
even in absence of sexual
reproduction. Since the evolution in the genotypic space is difficult
to analyze theoretically, we present in
Section~\ref{sec:PhenotypicSpeciation} a model of
phenotypic evolution which still exhibits the speciation effect and
that can be analyzed analytically. These approximate analytical
results are confirmed by Monte Carlo simulations in the effective
genotypic space, as reposted in Section~\ref{sec:Hypercubic}.

The conclusions are illustrated in the last Section. 

\section{Preliminaries}
\label{sec:Preliminarities}

\subsection{Genotypic space and mutations}
\label{sec:GenotypicSpace}
The fundamental ingredient of evolutionary  models is the genetic
information transmitted from one generation to the other, and the
selection mechanism that models the distribution of this information. 
In what follows we consider discrete generations,
since they are more directly implemented on a computer. 
The cellular automaton model that presented in Section~\ref{sec:CA}
is our microscopic reference system. By varying some parameters it may
present both non-overlapping and overlapping generations. 

In the simplest version, the genetic information ({\it genotype}) of an
individual is represented by a binary string $x=(x_1, x_2, \dots, x_L)$
of $L$ symbols $x_i=0,1$. In this way we are modeling haploid (only
one copy of each gene) organisms, i.e.\  bacteria or viruses or, more
appropriately, more archaic, pre-biotic entities. The choice of a binary
code is not fundamental but certainly makes things easier. It can be
justified by thinking of a purine-pyrimidine coding of DNA
 or of good and bad alleles for genes. In this second version, $0$
represents a good gene and $1$ a bad one. For bacteria, one can prove
that major and minor codons work in such a way~\cite{BagnoliLio}. 
Sometimes we  represent a genome using  Ising-like variables
$\sigma_i=2x_i-1$. 

During the individual life, errors (mutations) accumulate into the
germ line (which for unicellular organisms coincide with the somatic
line). When one individual reproduces, it duplicates its genetic material,
and during this process other mutations can occur. These errors are
transmitted to the progeny and cause the fact that even for asexual
organisms the offspring can be different from the parent.
We do not think that the differences between 
these two kinds of hereditary mutations
 causes qualitatively different behaviors,
at least at the level of schematization of this work. 

While there is a great variety of possible mutations (insertions,
deletions, transposition, inversions, etc),  from a mathematical
point of view it is preferable not to alter the length of the genome
(otherwise we should introduce a three-symbols code, say 1 for good
genes, -1 for bad genes and 0 for empty positions, i.e.\ neutral
genes). 

We  consider two kinds of mutations: point mutation, that interchange
a 0 with a 1, and all other mutation that do not alter the length of
the genome (transposition and inversions).

We can define the distance between two genomes $x$ and $y$
as the minimal number of point mutations needed to pass from $x$ to $y$,
and this coincides with the Hamming distance $d(x,y)$.
Thus, all possible genomes of length
$L$ are distributed on the $2^L$ vertices of an hypercube. 
A point mutation
corresponds to a unit displacement on that hypercube (short-range
jumps).

For simplicity we assume that
all point mutations are equally likely, while 
 in reality they depend on the identity of the symbol and 
on its positions on the genome. For real organisms, the probability of
observing a mutation is quite small. We assume that at most one
mutation is possible in one generation.
We denote with $\mu_s$ the probability
of having one point mutation per generation. 

The probability to have a point mutation from genotype $y$ to genotype $x$  
is given by
the short-range mutation matrix $\boldsymbol{M}_s(x,y)$ which is
\begin{equation}
\label{Ms}
	\boldsymbol{M}_s(x,y) = 
  \left\{\begin{array}{ll}
	  1-\mu_s & \mbox{if $x=y$,} \\
		  \dfrac{\mu_s}{L}& \mbox{if $d(x,y)=1$,} \\
	  0&\mbox{otherwise.}
  \end{array}\right.
\end{equation}
 
Other mutations correspond to long-range jumps in the genotypic space.
A very rough approximation consists in assuming all mutations
equiprobable. Let us denote with $\mu_{\ell}$ the probability per
generation of this kind of mutations. The long-range mutation matrix,
$\boldsymbol{M}_{\ell}$, is defined as
\begin{equation}
\label{Ml}
	\boldsymbol{M}_{\ell}(x,y) = 
	\left\{\begin{array}{ll}
		 1-\mu_{\ell} & \mbox{if $x=y$,} \\
		\dfrac{\mu_{\ell}}{2^L-1}&\mbox{otherwise.}
	\end{array}\right.
\end{equation}

In the real world, only a certain kind of mutations are possible, and in this
case  $\boldsymbol{M}_{\ell}$ becomes a sparse matrix $\hat{\boldsymbol{M}}_{\ell}$. We introduce a sparseness index
$s$ which is the average number of nonzero off-diagonal elements of
$\hat{\boldsymbol{M}}_{\ell}$. The sum of these off-diagonal elements 
still gives $\mu_{\ell}$. In this case $\hat{\boldsymbol{M}}_{\ell}$ 
is a quenched sparse
matrix, and $\boldsymbol{M}_{\ell}$ can be 
considered the average of the annealed version.

Both $\boldsymbol{M}_s$ and $\boldsymbol{M}_{\ell}$ are Markov matrices. 
Moreover, they are circular
matrices, since the value of a given element does not depend on its
absolute position but only on the distance from the diagonal. This
means that their spectrum is real, and that the largest
eigenvalue is $\lambda_0=1$. Since the matrices are irreducible, the
corresponding  eigenvector $\xi_0$ is non-degenerate, and corresponds
to the flat distribution $\xi_0(x) = 1/2^L$.
We discuss in details the properties of these matrices in Appendix~A.

\begin{figure}[b]
\centerline{\psfig{figure=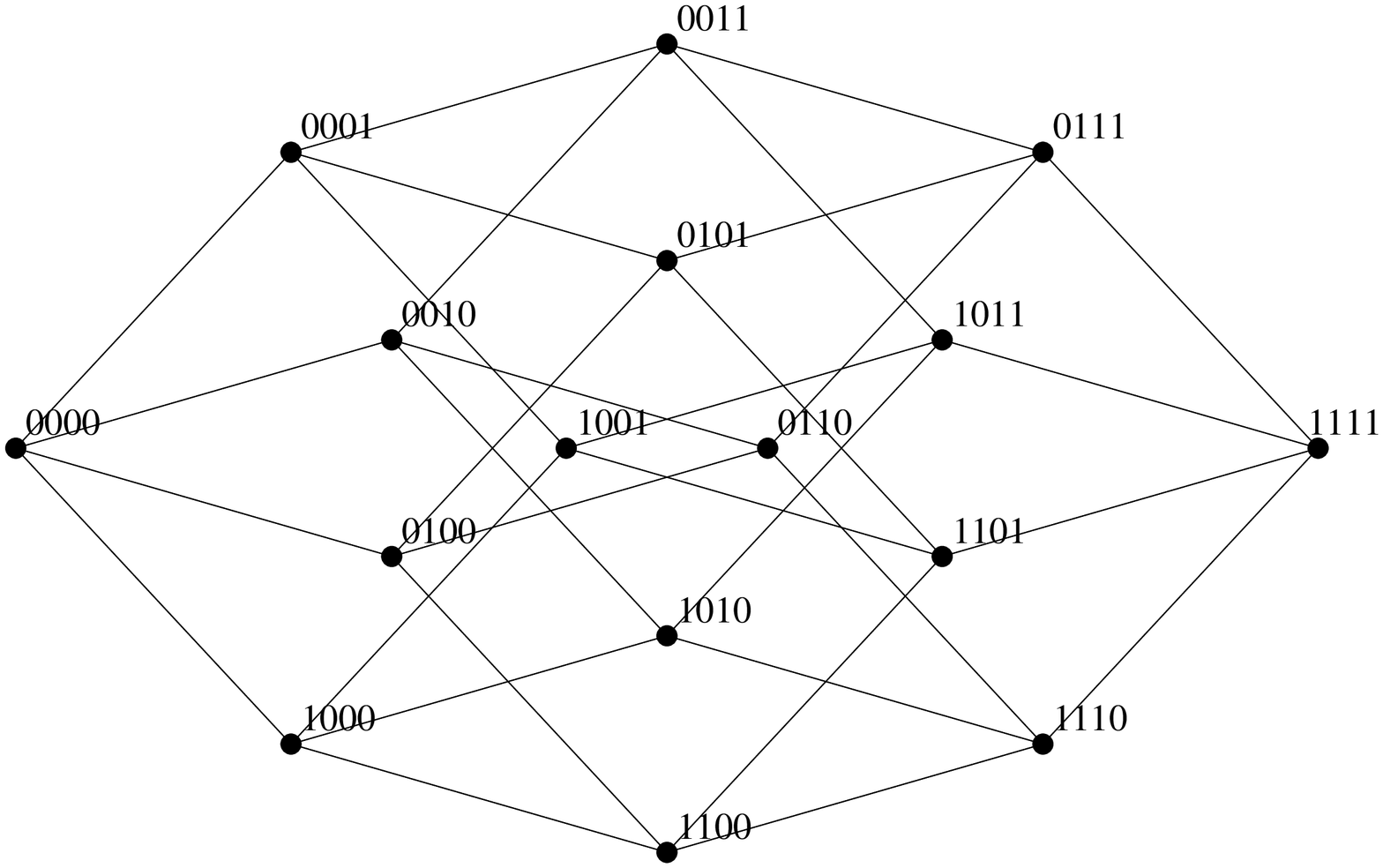,width=12cm,angle=270}}
\caption{\label{fig:Hyper} The easter egg representation of the Boolean hypercube for
$L=4$}
\end{figure}

We  use the following \emph{easter egg} representation for 
quasi-species  in the Boolean hyper-cubic space: 
starting from the origin of axis, we perform a step of  a fixed length
$R_0$ with an angle $n \pi/(2(L-1))-\pi/8$
if the $n$th bit ($0 \le n \le L$) of genotype $x$ has value one.    
In this way one locates the \emph{master sequence} (all zeros) 
 at the origin; the strains with one bad gene, distributed
according to the bad gene position
at distance $R_0$; the strains with two bad genes
at an approximate distance 
of $2 R_0$, and so on.
An example of the resulting
hypercube for $L=4$ is shown in Figure~\ref{fig:Hyper}.

\subsection{Phenotypic space and selection}
\label{sec:PhenotypicSpace}

The other necessary ingredient is the selection mechanism, which can
either modify the survival probability of individuals or 
their reproduction efficiency. 
The selection does not act directly on the genome, but rather on 
the {\it phenotype} (how an individual
appears to others). The phenotype $u(x)$ of a given genotype $x$ 
can be thought of as an array of
morphological characteristics, and generally lives in a much simpler
space than the genotype. We  consider the simplest case, for
which $u$ is just a real variable. Moreover we  consider $u$ as
a single-valued function of the genotype $u=u(x)$, which is not the
general case, since polymorphism or age dependence are usually
present.  

In general, the selection
mechanism is represented by the concept of the {\it fitness function}
$A$~\cite{Wright32,Hartle,Peliti95}, which
depends on the phenotype of the individual and on those present in the
 environment. The
fitness function
can be defined  as the average variation rate of the number of 
individuals that share that phenotype in a given environment,
\emph{if at least one individual were present}. 
This implies that
one can speak of  the fitness of phenotypes that never appeared in
the environment. 
The fitness function summarizes the effects of
reproduction and death rate, etc.
The more apt an
individual, the larger its fitness.

If direct interactions are not present, then the
fitness does not depend on the presence of other individuals,
 and it is a function of
the individual phenotype only. 
This leads to the concept of {\it fitness landscape}, analogous
to the potential surface for noninteracting particles. This 
analogy will be stressed in Section~\ref{sec:Path}. 
Evolution thus can be considered an {\it adaptive walk} in the fitness landscape
\cite{Kauffman87,Kauffman93} looking for the fitness optimum where
displacements in this space are due to mutations. 
In presence of a vanishing mutation rate and an infinite population, 
the system evolves until it reaches a local maximum of the fitness.
The effects of mutations and finite populations (genetic drift) can
lead to the escape from a local maximum (see Section~\ref{sec:Toy}).

In the presence of a flat part of the landscape
({\it neutral evolution}~\cite{Kimura}),
there is no preferred genotype, then the evolutive path is a random 
walk through the genotypic space~\cite{Derrida91, Higgs91}. 

As already noticed, the genotypic space is a high dimensional
hypercube. Analytical results in such a space are very difficult to obtain. 
The
phenotypic space is much simpler, and one is tempted to study
evolution in such a space. For instance,  assume that the phenotype
$u(x)$ is simply given by the sum of bits in $x$. The phenotypic space
is thus degenerate (there are several genotypes with the same phenotype)
and since point mutations change the phenotype by one unit, it is
possible to write the mean-field evolution in phenotypic space by
inserting the appropriate degeneration factors. For a generic fitness
function one can study the evolution in phenotypic space only in the
limit of a very flat landscape and a vanishing mutation rate,
assuming that mutations are able to connects all the genotypes, 
(for instance,  using long-range mutations).

We  use the following form for the phenotype $u$
\begin{equation}
\label{phenotype}
	u(x) = \dfrac{\mathcal{H}}{L} \sum_{i=1}^L \sigma_i + 
		\dfrac{\mathcal{J}}{L-1} \sum_{i=1}^{L-1}
		\sigma_i \sigma_{i+1} + \mathcal{K} \eta(x),
\end{equation}
where $\eta(x)$ is a random function of $x$, uniformly distributed between
$-1$ and $1$ ($\langle \eta(x) \eta(y) \rangle = \delta_{xy}$). 
The magnetic term $\mathcal{H}$ represents  a non-epistatic
 interaction among loci, $\mathcal{J}$ represents  a weakly
rough landscape, while $\mathcal{K}$ modulates a widely rough landscape (it can
be thought as an approximation of a spin-glass landscape).
 
We have not examined all possible combinations of all parameters. 
We set $\mathcal{J} = \mathcal{K} = 0$ except for the study in the almost flat
fitness landscape of Section~\ref{sec:Path}.

The exact definition of the fitness function is deferred to the next
section, in which a microscopic model will be presented. We  see
in Section~\ref{sec:MF} that in a mean field approach,  the 
reproduction rate can be approximated by
\begin{equation}
	A(u) = \exp\bigl(\beta H(u)\bigr), 
\end{equation}
which has a form reminiscent of a statistical mechanics model.  
We  call $H$ the fitness \emph{tout court}, while $\beta$ is a
parameter that can be used to modulate the steepness of $A$. We always
use $\beta=1$.
 
The fitness $H$
of the strain $x_i$ in the environment $\boldsymbol{x}=\{x^t_{1}, \dots,
x^t_{N}\}$ is defined as
\begin{equation}
 \label{fitnessH}
  H(x_i,\boldsymbol{x}) = V\bigl(u(x_i)\bigr) + 
   \dfrac{1}{N}\sum_{j=1}^{N} W\bigl(u(x_i),u(x_j)\bigr).
\end{equation}

The fitness is composed by two parts: 
the {\it static fitness} $V$, 
and the {\it interaction term}, whose kernel is the interaction matrix $W$.
The field $V$ represents a fixed or slowly changing environment;
the matrix
$W$  defines the chemistry of the world and is fixed in time.  
A phenotype $u$ with static fitness
$V(u)>0$ represents individuals that can survive in isolation
(say, after an inoculation into an empty substrate),  while a strain with
$V(u)<0$ represents predators or parasites 
that require the presence
of some other individuals to survive (again, the phenotype can also
not be present in the environment).
In this paper we generally consider a  static fitness function $V(u)$, 
with a single maximum.

The matrix $W$ mediates the interactions between two strains. 
For a classification in terms of usual
ecological interrelations, one has to consider together $W(u,v)$ and
$W(v,u)$.
One can have four cases:
\vspace{.3cm}
\begin{center}
\begin{tabular}{ccl}
$W(u,v)<0$ & $W(v,u)<0$ & competition \\
$W(u,v)>0$ & $W(v,u)<0$ & predation or parasitism of species $u$
on species $v$\\
$W(u,v)<0$ & $W(v,u)>0$ & predation or parasitism of species $v$
on species $u$\\
$W(u,v)>0$ & $W(v,u)>0$ & cooperation 
\end{tabular}
\end{center}
\vspace{.3cm}
  
The interaction matrix $W$ 
specifies the
sources of food for non autonomous strains, but due to the average over
the outside world, $W$ can be an arbitrary matrix.
Since the individuals with similar phenotypes are those sharing the largest
quantity of resources, the competition is the 
stronger the more similar their
phenotypes are (intra-species competition). 
This implies that the interaction matrix $W$ has negative components
on and near the diagonal. In this paper we only study this
feature, which we consider essential.  
 The interaction matrix $W$ assumes the following form:
\begin{equation}
  W(u,v) = - J  K\left(
			\dfrac{u-v}{R}\right)  
\label{W}
\end{equation}
where the parameter $J>0$   controls
the intensity  of the intra-species competition, and $R$ is the phenotypic range
over which the competition is effective.
 We  introduce a specific form for the competition
kernel $K((u-v)/R)$ in Section~\ref{sec:PhenotypicSpeciation}.    
 
\subsection{A cellular automata model for a simple ecosystem}
\label{sec:CA} 

We describe here a cellular automaton model that
should be considered as the microscopic reference model for the
following. We actually study only the spatial mean field
version of this model. 

Each individual occupies 
a cell of a lattice in a one dimensional space;
the size of the lattice is $Q$ sites. 
 This automaton has a large number of states, one for each different genome
plus a state ($*$) representing the empty cell. 
The evolution of the system is given by the
application of two rules: the {\bf survival} step, that includes 
the interactions among individuals,
and the {\bf reproduction} step.  

In the following,  a tilde labels quantities after
the survival step, and a prime those after the reproduction step.

\subsubsection{Survival}
\label{sec:survival}
 An individual $x_i\equiv x_i(t) \neq *$  
at time $t$ and site $i$, $i=1,\dots,Q$,  
has a probability $\pi$ 
of surviving per unit of time. It is reasonable to assume this probability 
to depend only on phenotypic characters. 
The survival probability $\pi=\pi(H)$ is expressed as a sigma-shaped function
of  the fitness function $A(H)$
\begin{equation}
\label{probability}
  \pi(H) = \dfrac{A(H)}{1+A(H)} = \dfrac{e^{\beta H}}{1+e^{\beta H}}=
     \dfrac{1}{2} + \dfrac{1}{2} \tanh(\beta H).
\end{equation}

The survival phase is  expressed as:
\begin{equation}
	\tilde x_i = 
	\left\{\begin{array}{ll}
			x_i & \mbox{with probability $\pi\Bigl(H\bigl(u(x_i),
			\boldsymbol{x}\bigr)\Bigr)$,} \\
			$*$ & \mbox{otherwise.}
	\end{array}\right.	
\end{equation}	
Clearly, empty cells remains empty
during survival. 

\subsubsection{Reproduction}
\label{sec:reproduction}
 Each non-empty cell tries to copy its state to all 
its $\mathcal Q$ neighboring 
cells with probability $P_r$. If two or more cells try to copy themselves 
in the same site, the conflict is solved choosing one of them randomly.
We are interested in two limiting cases: 
\begin{description}
\item[(a)] $P_r=1/{\mathcal Q}$:  an individual
 colonizes in average one of all 
its neighboring cells.
\item[(b)] $P_r=1$:  an individual colonizes all its neighboring cells.
\end{description}

These rules can be implemented more easily as
rules  for empty cells.  In case {\bf (a)} each empty cell
chooses randomly one of the  neighboring cells (either empty or non-empty)
and copies its state.  If the newborn state  is different from the
empty state then  mutations apply by reversing the value of one bit
with probability $\mu$. This implementation is completely equivalent
to rule {\bf (a)} for long-range couplings.  This case, 
due to the low value of
$P_r$,  is particularly suitable to study  the conditions
that lead to the extinction of the whole population. 

In case $\bf (b)$ each empty cell chooses randomly one of the
neighboring non-empty cells  and copies its state; then mutations apply 
by reversing the value of one bit with probability $\mu$.   
In the spatial mean-field approximation, the neighborhood extends 
to the whole population
(${\mathcal Q} \rightarrow
Q$); in this limit, rule $\bf (b)$  is characterized by a constant population.

In both cases we can notice that  the effective reproduction rate
does not only depend  on the survival probability of the individual,
but also on total availability of empty cells.

\subsection{Spatial mean field theory and the selection mutation
equation} 
\label{sec:MF}

One can have an insight of the features of the model by means of a
simple spatial mean field analysis. Let $n(x)\equiv n(x,t)$ be  the
number of organisms with genotype $x$, $N\equiv N(t)$ is the total number
of occupied cells
\[
	N= \sum_x n(x),
\]
$m = N/Q$ is the average fraction of occupied cells
and $n_*$ the number of empty
sites, with $n_*+N = Q$.

Let us consider the two reproduction rules separately.

\subsubsection{Case (a): Variable Populations}
We can express the fitness $H$ (and thus the survival
probability $\pi$) in terms of the number of individuals $n(x)$ 
in a given strain or in terms of the probability distribution $p(x)=n(x)/N$
\begin{eqnarray}
	H(u(x),\boldsymbol{n}) &=& V(u(x))+\dfrac{1}{Q}\sum_y
		W\bigl(u(x),u(y)\bigr)n(y);\nonumber\\
	H(u(x),\boldsymbol{p})&=& V(u(x))+m\sum_y W\bigl(u(x),u(y)\bigr)p(y) .
		\label{Hp}
\end{eqnarray}
where $\boldsymbol{p}$ denotes the probability distribution.
The average evolution of the system  will be governed
by the following equations:
\begin{equation}
\begin{array}{rl}
	\tilde n(x) &=\pi\bigl(u(x),\boldsymbol{n}\bigr)n(x), \\
	n^{\prime }(x) &=\tilde n(x)+\dfrac{\tilde n_*}Q\sum_yM(x,y)\tilde n(y). 
\end{array}
\label{n}
\end{equation}

Using the Markov properties of the mutation matrices $M$,
and summing over $x$ in Eqs.~(\ref{n}),
 we obtain an equation for $m$:
\[
	\begin{array}{rl}
	\widetilde{m} &=\dfrac{\sum_x\tilde n(x)}{Q}=
	 \dfrac{1}{Q}\sum_x\pi\bigl(u(x),\boldsymbol{n}\bigr)n(x)=
	 m\overline{\pi} \\
	m^{\prime } &=\dfrac{\sum_xn^{\prime }(x)}{Q}=
	 \widetilde{m}+\dfrac{\tilde n_*}{%
Q^2}\sum_{y}\tilde n(y), 
	\end{array}
\]
i.e.
\begin{equation}
 	m'=m\overline{\pi}(2-m\overline{\pi}), \label{logistic}
\end{equation}
where
\[
	\overline{\pi}\equiv \dfrac{1}{N}
	\sum_x\pi\bigl(u(x),\boldsymbol{n}\bigr)n(x)=
		\sum_x\pi\bigl(u(x),\boldsymbol{p}\bigr)p(x)
\]
is the average survival probability.

The normalized evolution equation for $p(x)$ is:
\begin{equation}
	p^{\prime }(x)=\dfrac{
	\pi\bigl(u(x),\boldsymbol{p},m\bigr)p(x)+(1-m\overline{\pi})\sum_yM(x,y)
		\pi\bigl(u(y),\boldsymbol{p},m\bigr)p(y)}
	{\overline{\pi}(2-m\overline{\pi})}.   \label{peq}
\end{equation}

Notice that Eq.~({\ref{logistic}}) is a logistic equation with $\overline{\pi}$ 
as control parameter.  The stationary condition, ($m'=m$), is
\begin{equation}
	m =\dfrac{2\overline{\pi}-1}{\overline{\pi}^2}.
	\label{mequation}
\end{equation}

One observes extinction if $\overline{\pi} \le 1/2$. 
The decrease of $\overline{\pi}$ can be induced by a  variation of the
environment (notably $V(x)$) or by an increase of the mutation rate
$\mu$, which broadens the distribution $p(x)$.
This last effect 
corresponds to
the mutational meltdown phenomenon, for which the population vanishes while
continuing to exhibit a quasi-species distribution (see
Section~\ref{sec:Toy}). 
Since the total
population $m$ multiplies the competition term in Eq.~(\ref{peq}), one
cannot observe coexistence of species due to competition near the
mutational meltdown transition. 
  From Eq.~({\ref{logistic}})
one could expect a periodic
or chaotic behavior of the population; 
however, since $\overline{\pi}$ is always less than one, the asymptotic
dynamics of the 
population $m$ can only exhibit fixed points. 

\subsubsection{Case (b): Constant Populations}
\label{const}

In the spatial mean-field limit  (${\mathcal Q} \rightarrow Q$), 
the size of the whole populations does not change after 
survival and reproduction.  Because all the empty sites are populated by new
individuals (except the trivial case of all empty sites
as initial condition) $N=Q$. In this case
we can rewrite the fitness function $H$ as function of $p(x)=n(x)/N$

\[
	\begin{array}{rl}
		H(u(x),\boldsymbol{n}) &= V(u(x))+\dfrac{1}{N}\sum_yW(u(x),u(y))n(y)\\
		H(u(x),\boldsymbol{p})&= V(u(x))+\sum_yW(u(x),u(y))p(y).
	\end{array}
\]

The evolution equations after the survival (labeled by a tilde as
before) and reproduction (labeled by a prime) are

\[ 
	\begin{array}{rl}
		\tilde n(x) &=\pi\bigl(u(x),\boldsymbol{n}\bigr)n(x),  \label{fixedtilde} \\
		n^{\prime }(x) &=\tilde n(x)+\dfrac{\tilde n_*}{\tilde N}\sum_yM(x,y)\tilde n(y). 
		\label{fixedprime}
	\end{array}
\]

 The normalized evolution equation for $p(x)$ is
\begin{equation}
	p'(x)=
	\pi\bigl(u(x),\boldsymbol{p}\bigr)p(x)+\left(\dfrac{1}{\overline{\pi}}-1\right)
	\sum_yM(x,y)\pi\bigl(u(y),\boldsymbol{p}\bigr)p(y),   \label{fixedpeq}
\end{equation} 
where the average survival probability $\overline{\pi}$ is defined as 
\[
	\overline{\pi}\equiv 
		\dfrac{1}{N}\sum_x\pi\bigl(u(x),\boldsymbol{n}\bigr)n(x)=
		\sum_x\pi\bigl(u(x),\boldsymbol{p}\bigr)p(x)
\]

Since the size of whole population is constant,  the mutational
meltdown effect is not present, 
and we can consider the limit of small survival probability
(non-overlapping generations) $\pi \rightarrow 0$. In this limit
$\pi(u(x),\boldsymbol{p}) \simeq A(x,\boldsymbol{p})$ and Eq.~(\ref{fixedpeq}) becomes
\begin{equation}
	p^{\prime }(x)=
	\dfrac{
	\sum_y M(x,y)A\bigl(u(y),\boldsymbol{p}\bigr)p(y)}{\overline A}.  
	 \label{mutsel}
\end{equation} 
with 
\[
{\overline A}= \sum_x A\bigl(u(x),\boldsymbol{p}\bigr) p(x)
\]
 
Eq.~(\ref{mutsel}) is known in the literature~\cite{mutsel} as the
mutation selection equation, and defines a reaction-diffusion
process in the genotypic space. An equivalent form is
\begin{equation}
	p'(x)=
	\dfrac{
	A\bigl(u(x),\boldsymbol{p}\bigr)\sum_y M(x,y)p(y)}{\overline A}. \label{mutsel1}
\end{equation} 
obtained from Eq.~(\ref{mutsel}) by a shift of an half time step.

For a vanishing mutation rate (i.e.\ only assuming that all phenotypes
can be generated), 
the average fitness $\overline{A}$ of Eq.~(\ref{mutsel})
is a monotonically increasing
function of time~\cite{Fisher,Peliti:review}.

\section{Evolution on a static fitness landscape}

\subsection{Path integral formulation of evolving ecosystems}
\label{sec:Path}

Let us consider the case in which the fitness $A$ does not depend
explicitly on the whole distribution $\boldsymbol{p}$, i.e. no
competition.
Eq.~(\ref{mutsel}) can be linearized by using unnormalized variables 
$z(x,t)$
that satisfy
\begin{equation}
	z(x, t+1) = \sum_{y} A(u(y)) M(x,y)z(y,t), \label{z}
\end{equation}
with the correspondence
\[
	p(x,t) = \dfrac{z(x,t)}{\sum_{y}z(y,t)}.
\]
In vectorial terms, Eq.~(\ref{z}) can be written as 
\begin{equation}
	\boldsymbol{z}(t+1) = \boldsymbol{M}\boldsymbol{A}\boldsymbol{z}(t),
	\label{Z}
\end{equation}
where $\boldsymbol{M}_{xy}=M(x,y)$ and $\boldsymbol{A}_x=A(u(x))\delta_{xy}$.

When one takes into consideration only point mutations ($\boldsymbol{M}\equiv \boldsymbol{M}_s$),
Eq.~(\ref{z}) can be read as 
the transfer matrix of a two-dimensional Ising
model~\cite{Leuthausser,Tarazona,Baake}, 
for which the genotypic element $\sigma_i^t$ corresponds to the spin in
row $t$ and column $i$, and $ z(\boldsymbol{\sigma},t)$ is the restricted partition function of
row $t$.  The effective Hamiltonian (up to  constant terms) 
of a possible genealogical story $\{\boldsymbol{\sigma}^t\}$ from time $1\le t \le T$ is
\begin{equation}
	\mathcal{V} = \sum_{t=1}^{T-1} \left(\gamma \sum_{i=1}^L
	\sigma_i(t)\sigma_{i}(t+1) + V\bigl(u(x)(t)\bigr)\right),
	\label{ising} 
\end{equation}
where $\gamma=-\ln(\mu_s/(1-\mu_s))$.
 
This peculiar two-dimensional Ising model has a long-range coupling along
the row (depending on the choice of the fitness function) and a
ferromagnetic coupling along the time direction (for small short
range mutation  probability).  In order to obtain the statistical
properties of the system one has to sum over all possible
configurations (stories), eventually selecting the right boundary
conditions at time $t=1$. 

The bulk properties of Eq.~(\ref{ising})  cannot be reduced in general
to the equilibrium distribution of a one
dimensional system, since the transition probabilities among rows do
not obey detailed balance. Moreover, the temperature-dependent
Hamiltonian (\ref{ising}) does not allow an easy identification
between energy and selection, and temperature and mutation, what is
naively expected by the biological analogy with an adaptive walk. 

An Ising configuration of Eq.~(\ref{ising}) corresponds to a possible
genealogic story, i.e. as a directed polymer in the genotypic 
space~\cite{Galluccio}, where mutations play the role of elasticity.
It is natural to try to rewrite the model in terms of the sum over all
possible paths in genotypic space. 

\subsubsection{Long-range mutations}.
Let us first consider the long-range mutation case. 

Eq.~(\ref{Z}), reformulated according to Eq.~(\ref{mutsel1}), corresponds to 
\[
	\boldsymbol{z}(t+\tau) = (\boldsymbol{A}\boldsymbol{M}_{\ell})^\tau\boldsymbol{z}(t).
\]
Since it is easier to consider the effects of 
$\boldsymbol{A}$ and $\boldsymbol{M}_{\ell}$ separately, let us 
study in which limit they commute.  

The norm of the commutator on the asymptotic probability distribution
$\boldsymbol{p}$ is 
\[
||[\boldsymbol{A}\boldsymbol{M}_{\ell}]||=\sum_{ij}
|[\boldsymbol{A}\boldsymbol{M}_{\ell}]_{ij}\boldsymbol{p}_j|,
\]
and it is bounded by $\mu_{\ell}c$, where $c= \max_{ij}
|A_{ii}-A_{jj}|$.
In the limit $\mu_{\ell}c\rightarrow 0$ (i.e.\ a very smooth landscape), 
to first order in $c$ we have
\[
	(\boldsymbol{A}\boldsymbol{M}_{\ell})^\tau = \boldsymbol{A}^\tau\boldsymbol{M}_{\ell}^\tau +
	O(\tau^2 \mu_{\ell} c) \boldsymbol{A}^{\tau-1}\boldsymbol{M}_{\ell}^{\tau-1}, 
\]
which is the analogous of the Trotter product formula. 

When $\tau$ is order $1/\mu_{\ell}$, $\boldsymbol{M}_{\ell}^\tau$ is a constant
matrix with elements equal to $1/2^L$, and thus $\boldsymbol{M}_{\ell}\boldsymbol{p}$ is a 
constant probability distribution. If $\mu_{\ell}$ is large enough, 
$(\boldsymbol{A}\boldsymbol{M}_{\ell})^\tau = \boldsymbol{A}^\tau\boldsymbol{M}_{\ell}^\tau$.

The asymptotic probability distribution  
$\tilde{\boldsymbol{p}}$
is thus proportional to the diagonal of $\boldsymbol{A}^{1/\mu_{\ell}}$:
\begin{equation}
	\tilde p(x) = C \exp\left(\dfrac{V(u(x))}{\mu_{\ell}}\right)
	\label{Boltzmann}
\end{equation}
i.e.\ a Boltzmann distribution with Hamiltonian $V(u(x))$ 
and temperature $\mu_{\ell}$. This corresponds to the naive analogy between
evolution and equilibrium statistical mechanics. In other words, the
genotypic distribution is equally populated if the phenotype is the
same, regardless of the genetic distance since we used long-range
mutations.

In terms of the sum over paths~\cite{Schulman,Bagnoli:path},
Eq.~(\ref{z}) becomes
\[
	z(\boldsymbol{\sigma}, T) = \sum_{\boldsymbol{\sigma}'(T-1)} \dots \sum_{\boldsymbol{\sigma}'(0)} 
		\exp\left(\sum_{t=0}^{T-1} -\gamma d_L(\boldsymbol{\sigma}'(t+1), \boldsymbol{\sigma}'(t))
		-V(\boldsymbol{\sigma}'(t)\right) z(\boldsymbol{\sigma}'_0, 0)
\]
where $\boldsymbol{\sigma}'_0=\boldsymbol{\sigma}'(0)$ and $\boldsymbol{\sigma}=S'(T)$. In terms of directed polymers 
$\gamma d_L(\boldsymbol{\sigma}'(t+1), \boldsymbol{\sigma}'(t))$ is the bending contribution to the energy, 
while in term of path sums $\gamma d_L(\boldsymbol{\sigma}'(t+1), \boldsymbol{\sigma}'(t))$ is the kinetic energy 
(and $\gamma$ is the mass of the particle).

The relevant paths in the mean field spirit are those that insist on a 
given genotype for a time order $1/\mu_{\ell}$. Restricting the sum to the
paths formed by segments of time length $1/\mu_{\ell}$, 
the kinetic energy contributes a constant term. After rescaling the time 
of a factor $\tau = 1/\mu_{\ell}$, we have a free sum of the type 
\[
	z(\boldsymbol{\sigma}, n \tau) = \sum_{[ \boldsymbol{\sigma}'(n) ]}  
		\exp\left(\sum_{i=0}^{n-1} 
		\tau V\bigl(\boldsymbol{\sigma}'(i)\bigr)\right) z(\boldsymbol{\sigma}'_0, 0)
\]
which gives the Boltzmann probability distribution~(\ref{Boltzmann}).

We have checked numerically this mapping using a flat landscape
$V(u)=u$ and the complete phenotype~(\ref{phenotype}).
We iterated Eq.~(\ref{mutsel}) for a given genome length $L$ and for a time
$t$ large enough to be sure of having reached the asymptotic state. 
We plotted the logarithm of the probability
distribution $\tilde p(x)$ versus the value of the Hamiltonian
$V(u(x))$. We computed the slope $1/\mu_{e}$ and the average of
quadratic differences  $\chi^2$ of the linear regression. 

\begin{figure}
	\begin{tabular}{cc}
	 \psfig{figure=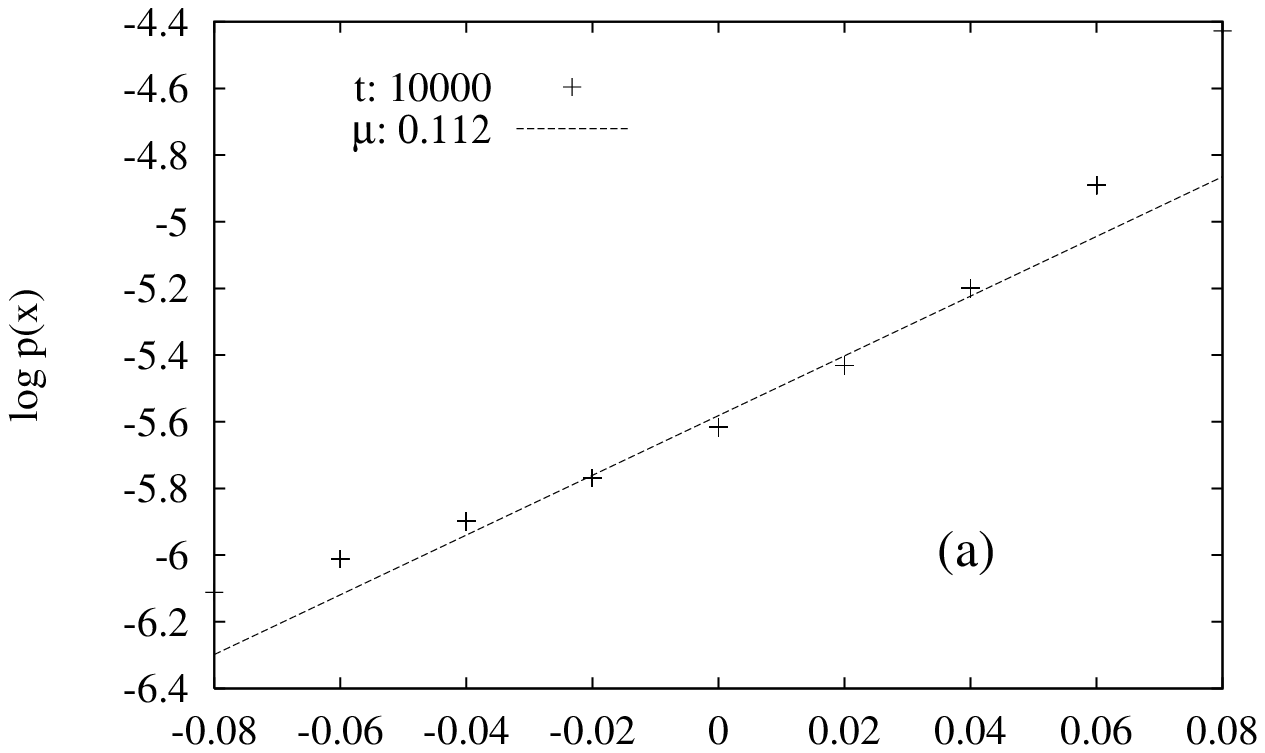,width=8cm} &
	\psfig{figure=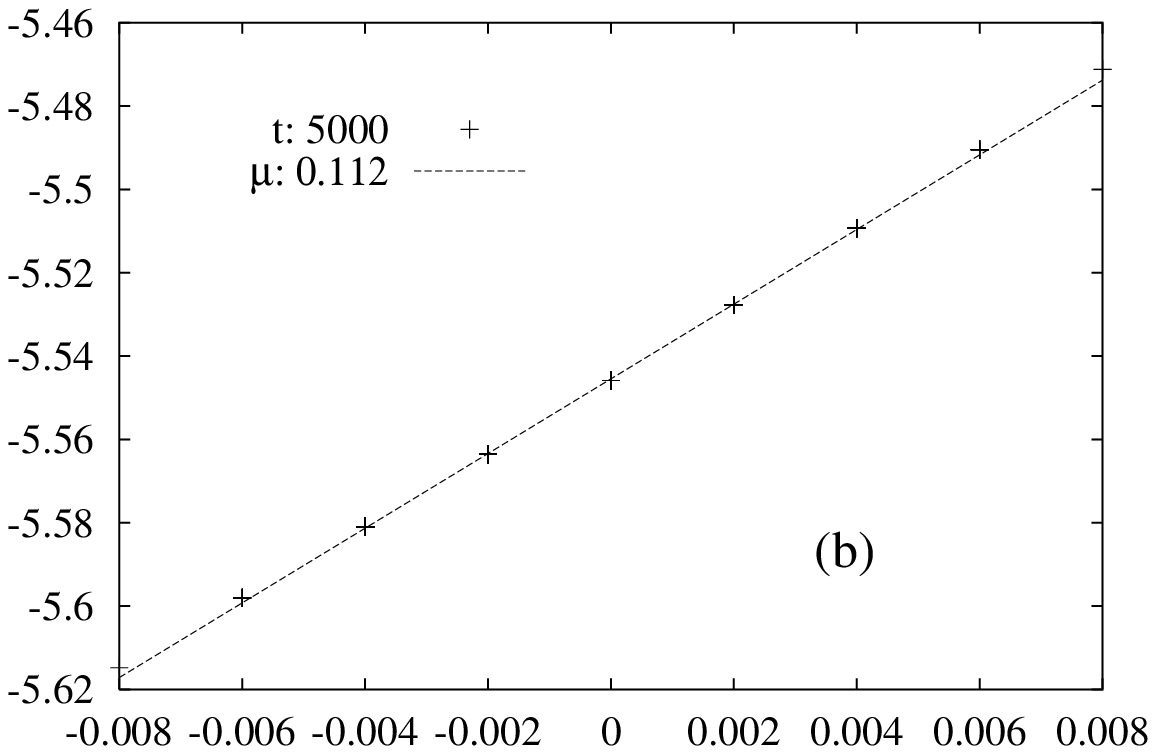,width=8cm}\\
	 \psfig{figure=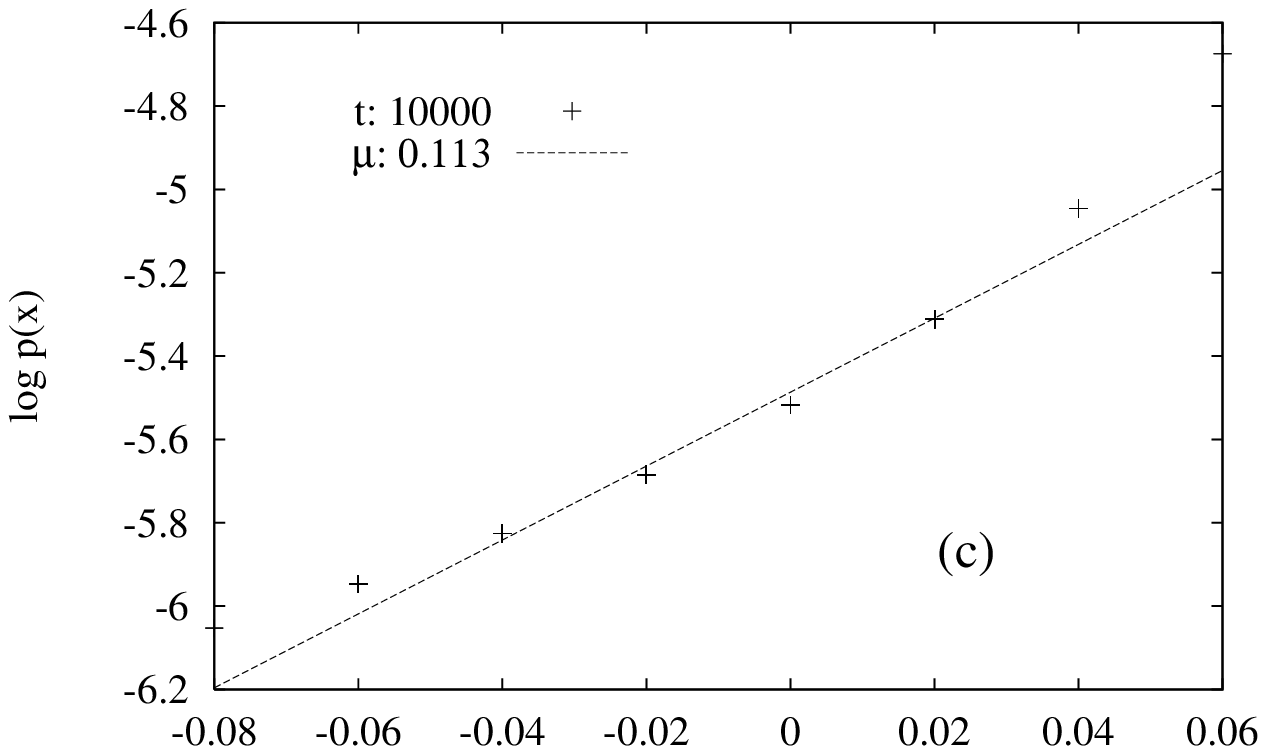,width=8cm} & 	
	 \psfig{figure=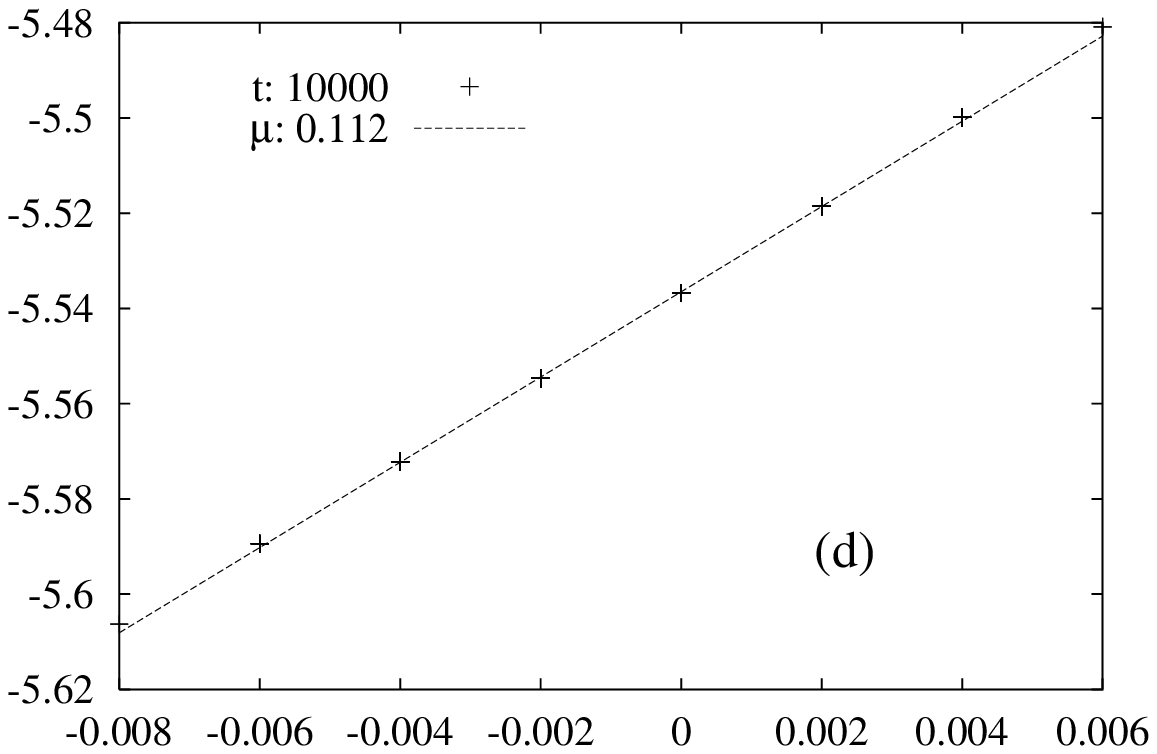,width=8cm} \\
	 \psfig{figure=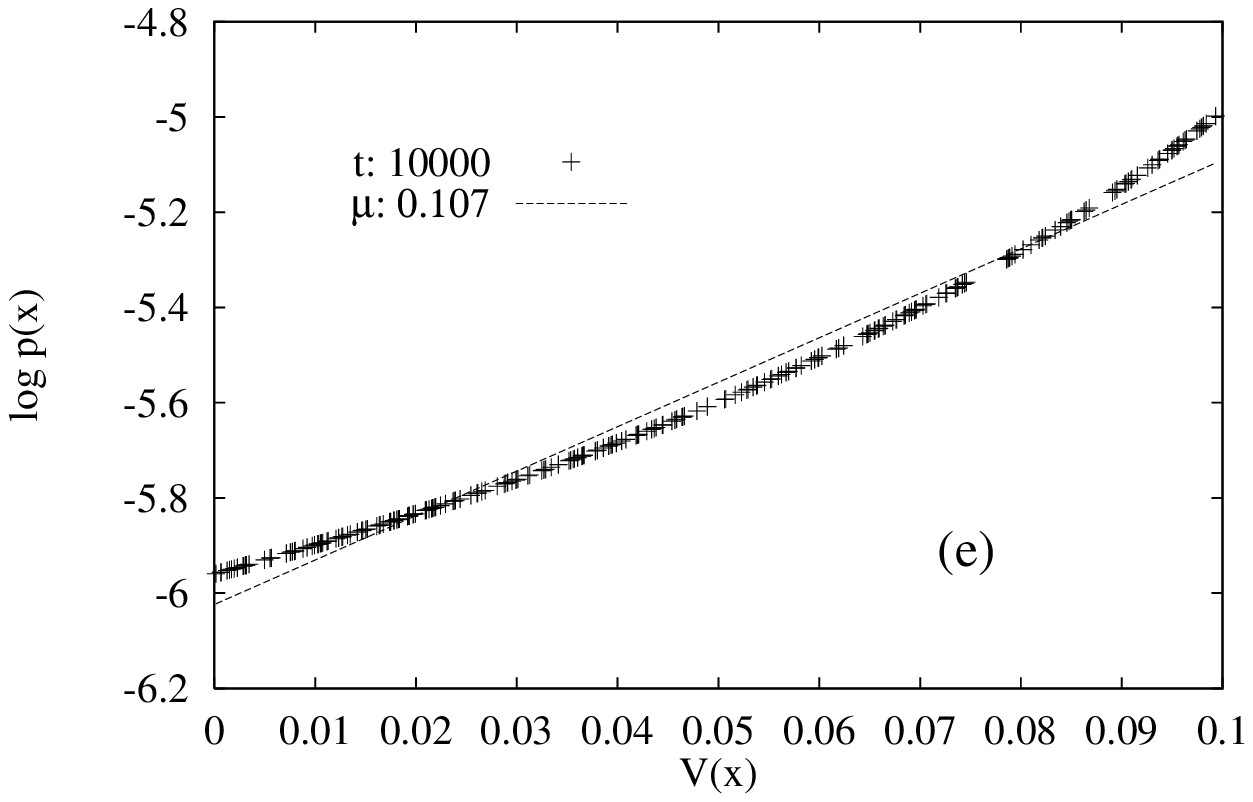,width=8cm} & 	
	 \psfig{figure=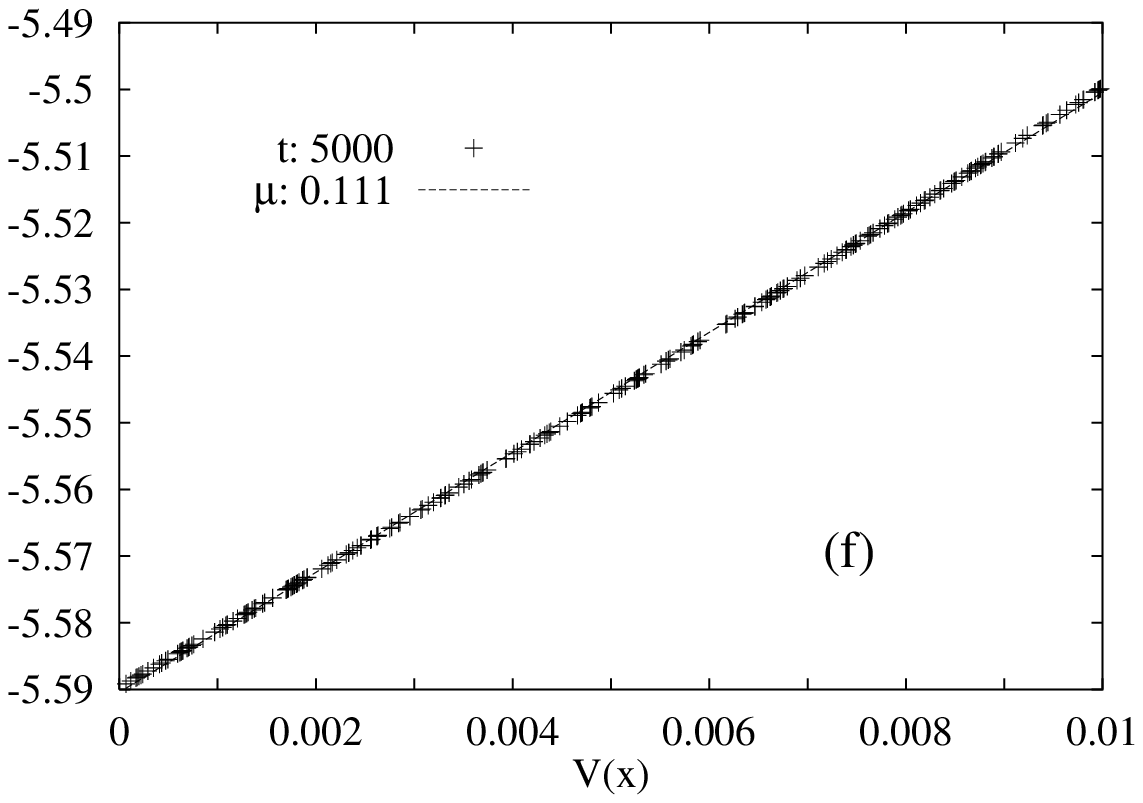,width=8cm} 
	\end{tabular}
	\caption{
	Numerical check for long-range mutations. In the simulations we set
	$L=8$, $\mu_{\ell}=0.1$ and  $\mu_s=0$.
	We varied  $\mathcal{H}$ (a,b), $\mathcal{J}$ (c,d) and $\mathcal{K}$ (e,f),
	setting all other parameters to zero.
	(a)  $\mathcal{H}=0.01$, $\mathcal{J}=0$, $\mathcal{K}=0$;
	(b)  $\mathcal{H}=0.001$, $\mathcal{J}=0$, $\mathcal{K}=0$;
	(c) $\mathcal{H}=0$, $\mathcal{J}=0.01$, $\mathcal{K}=0$;
	(d) $\mathcal{H}=0$, $\mathcal{J}=0.001$, $\mathcal{K}=0$;
	(e) $\mathcal{H}=0$, $\mathcal{J}=0$, $\mathcal{K}=0.1$;  
	(f) $\mathcal{H}=0$, $\mathcal{J}=0$, $\mathcal{K}=0.01$. In the figures $t$
	indicates the number of generations, $\mu$ the reciprocal of the slope of
	linear regression ($\mu_{e}$ in the text).   
	}
	\label{figure:Ml}
\end{figure}

The results are shown in
Figure~\ref{figure:Ml}.
We see that the equilibrium hypothesis is well verified in the limit
$\mu_{\ell} \gg c$; and that convergence is faster for a rough landscape. 

\begin{figure}
	\psfig{figure=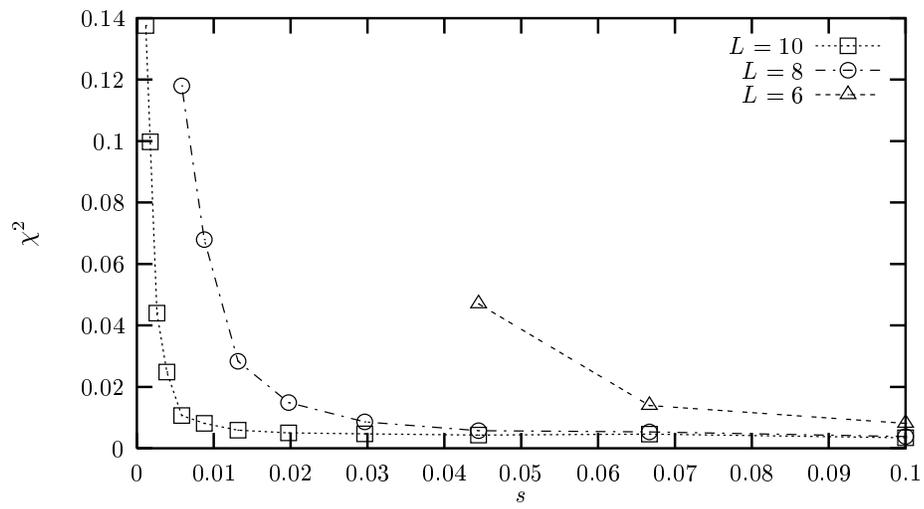,width=12cm}
	\caption{Scaling of $\chi^2$ with the sparseness factor $s$, for
	three values of genome length $L$.} 
	\label{figure:sparsechi} 
\end{figure}

Since in reality the long-range mutations follow preferred patterns,
we checked that the inclusion of a sparseness factor $s$ (the number
of off-diagonal elements different from zero) do not alter
these results. In Figure~\ref{figure:sparsechi} we show that the
average of quadratic differences $\chi^2$ keeps 
low even for very small values of the sparseness $s$, and that the transition
point vanishes when increasing $L$. 
The slope of the line is
almost constant for all values of the sparseness factor $s$.  This implies that
for large enough genomes the sparseness of the long range mutation matrix does
not alter the statistical mechanics interpretation of selection and mutations.

\subsubsection{Short-range mutations}

\begin{figure}
	\begin{tabular}{cc}
	\psfig{figure=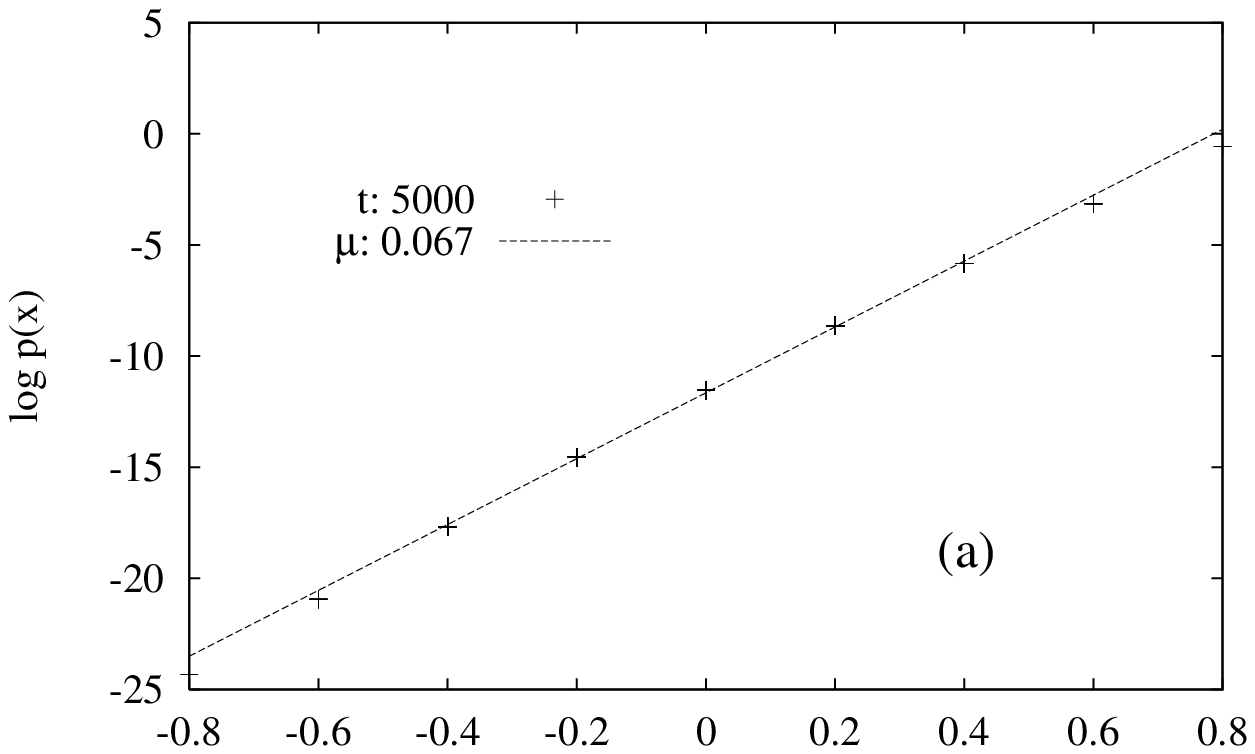,width=7cm} & 	
	\psfig{figure=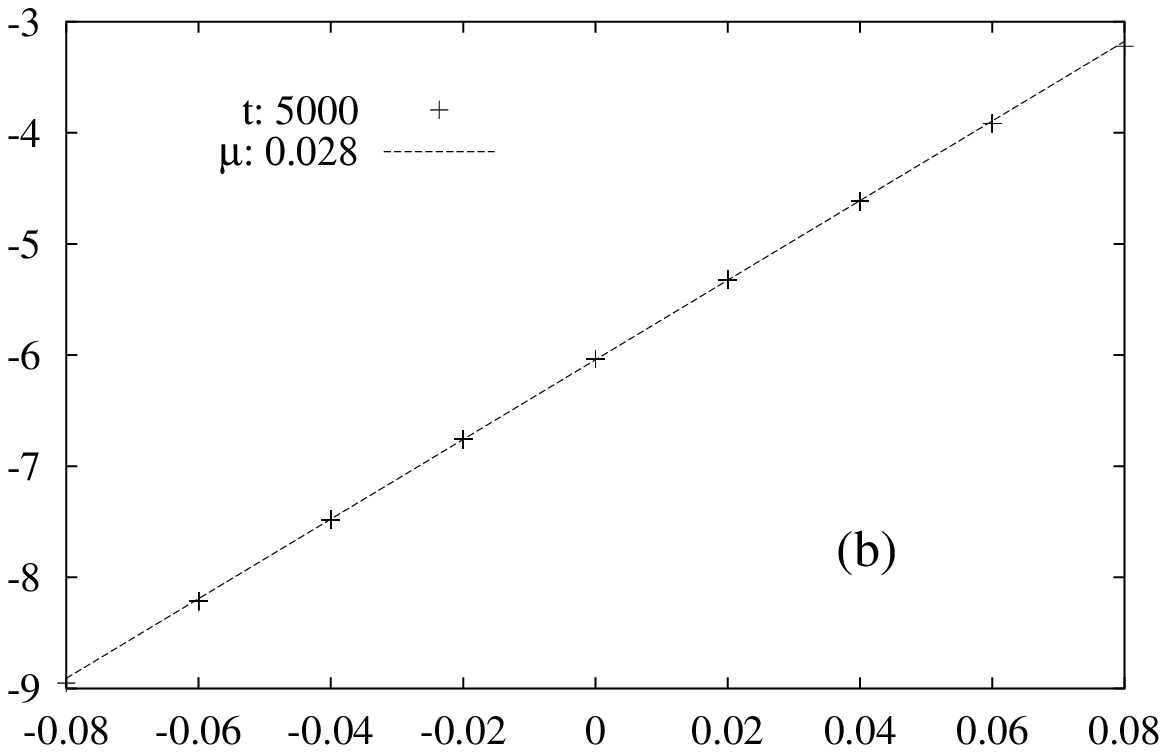,width=7cm} \\
	\psfig{figure=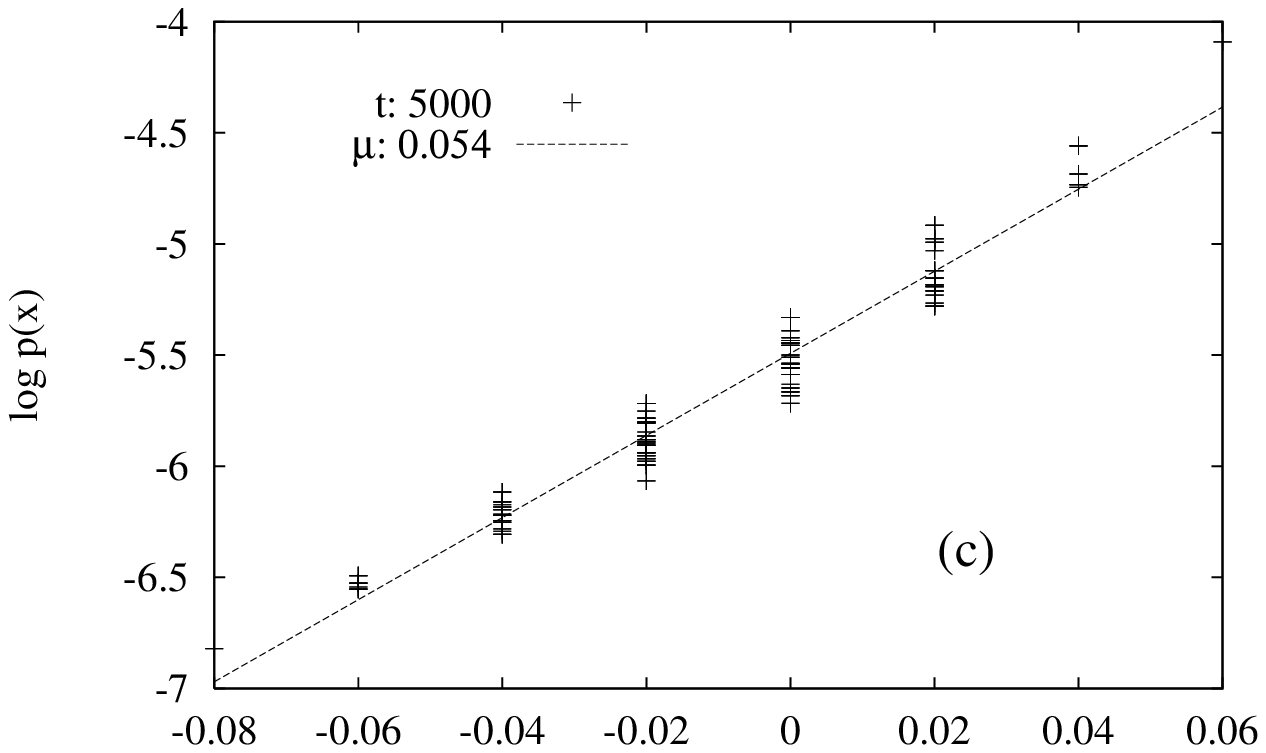,width=7cm} & 	
	\psfig{figure=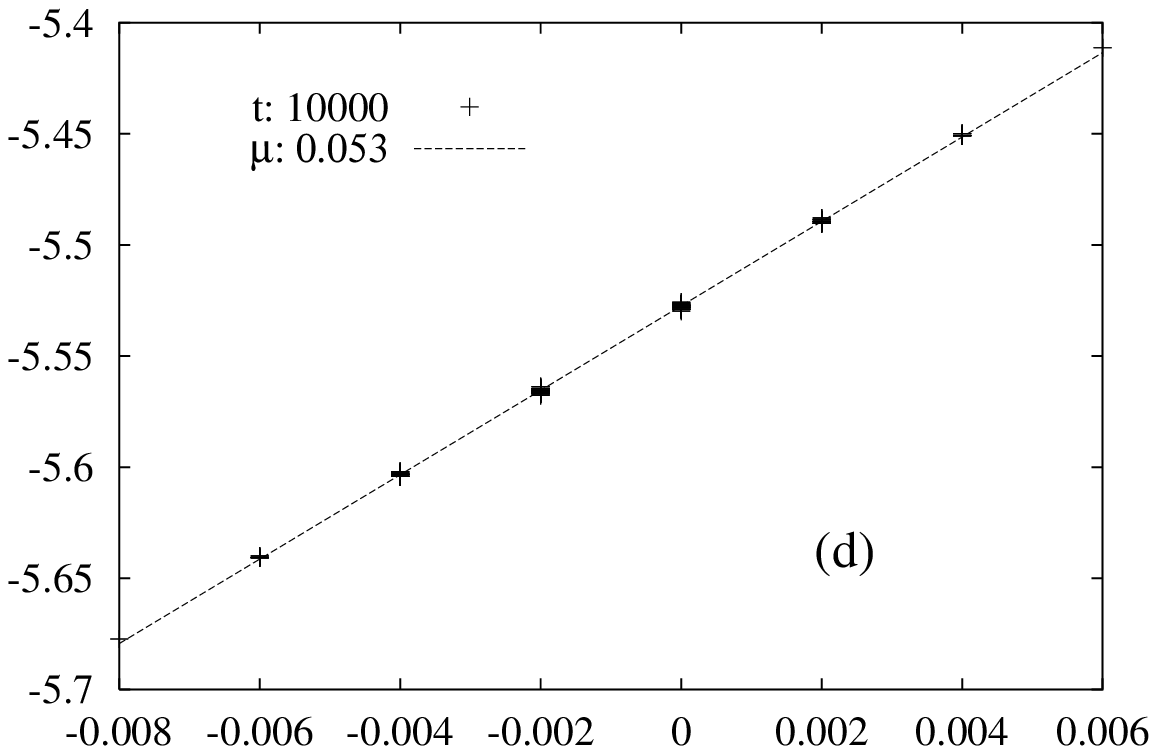,width=7cm} \\
	\psfig{figure=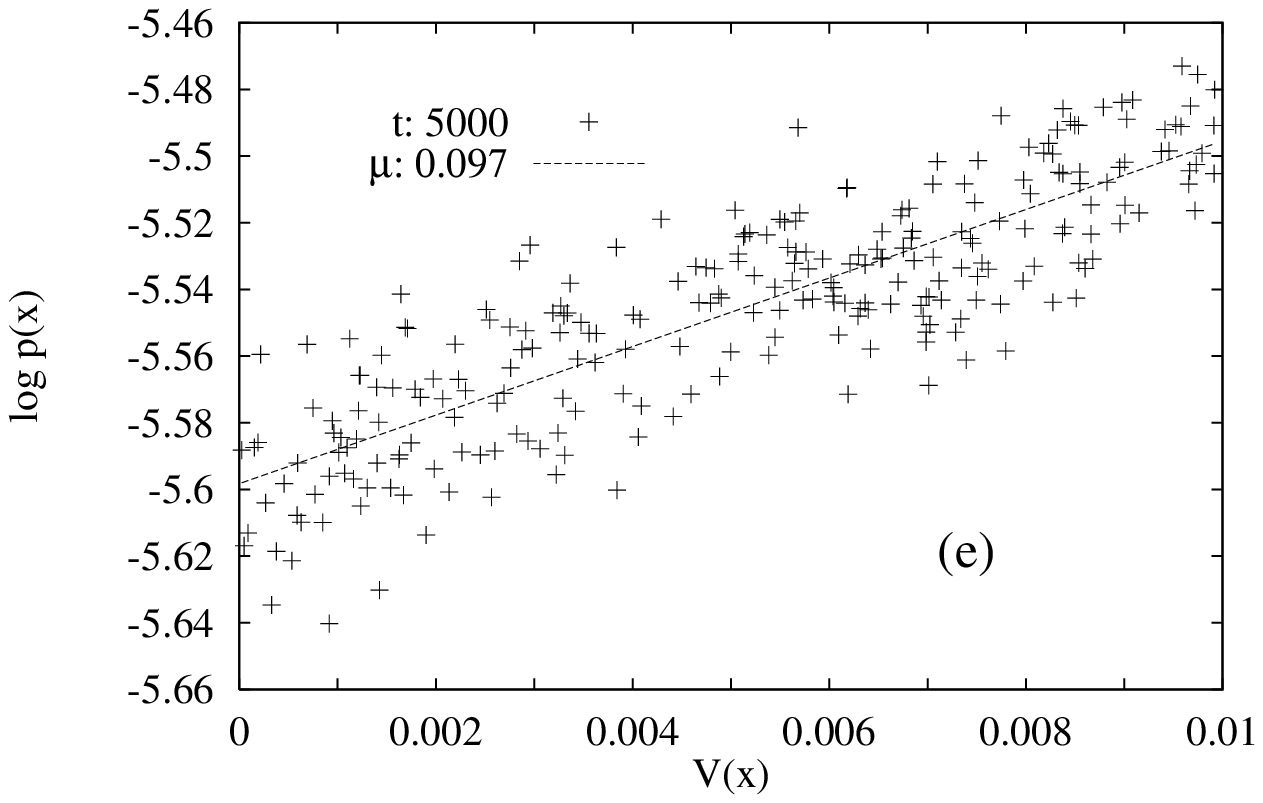,width=7cm} &
	\psfig{figure=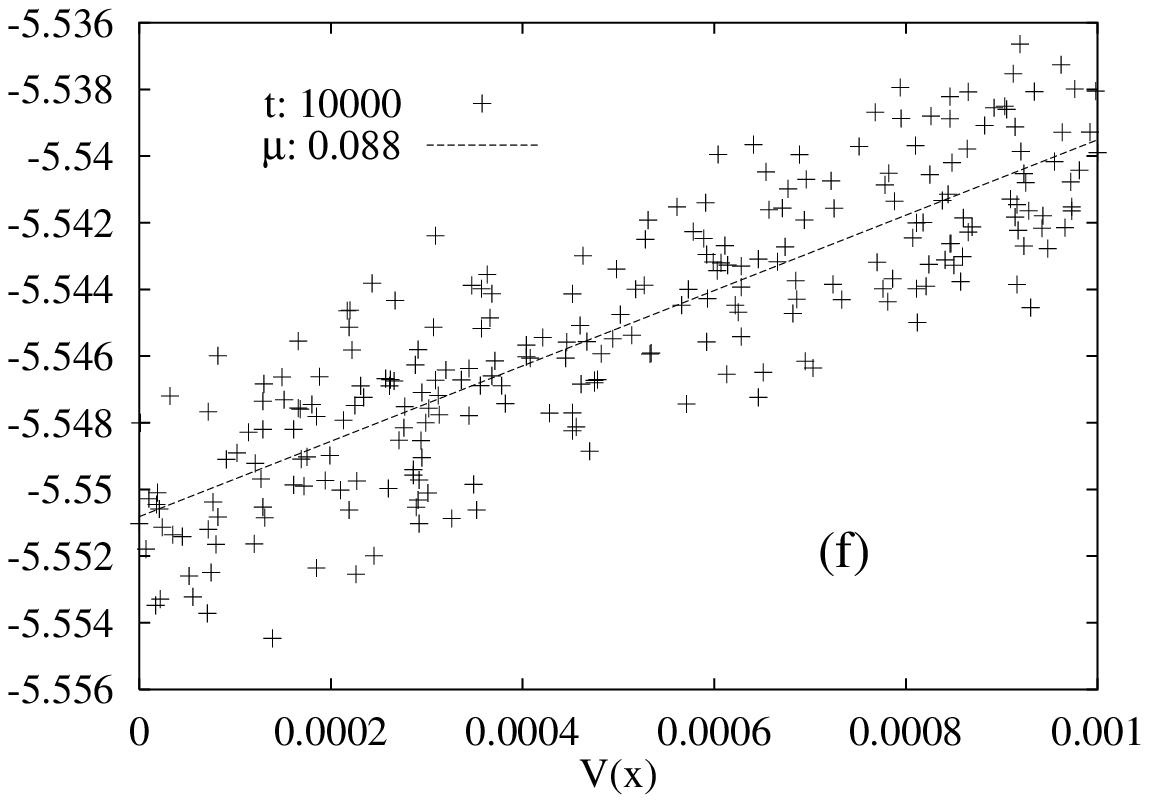,width=7cm}
	\end{tabular}
	\caption{
	Numerical check for short-range mutations. In the simulations we set
	$L=8$, $\mu_{\ell}=0$ and  $\mu_s=0.1$.
	We varied  $\mathcal{H}$ (a,b), $\mathcal{J}$ (c,d) and $\mathcal{K}$ (e,f),
	setting all other parameters to zero.
	(a)  $\mathcal{H}=0.1$, $\mathcal{J}=0$, $\mathcal{K}=0$;
	(b)  $\mathcal{H}=0.01$, $\mathcal{J}=0$, $\mathcal{K}=0$;
	(c) $\mathcal{H}=0$,  $\mathcal{J}=0.01$,$\mathcal{K}=0$;
	(d) $\mathcal{H}=0$, $\mathcal{J}=0.001$, $\mathcal{K}=0$;
	(e) $\mathcal{H}=0$, $\mathcal{J}=0$, $\mathcal{K}=0.01$;  
	(f) $\mathcal{H}=0$, $\mathcal{J}=0$, $\mathcal{K}=0.001$. 
	In the figures $t$
	indicates the number of generations, $\mu$ the reciprocal of the slope of
	linear regression ($\mu_{e}$ in the text).   
	}
	\label{figure:Ms}
\end{figure}

The above results hold qualitatively also for short-range mutations
as shown in Figure~\ref{figure:Ms}. A small dispersion of points in
Figure implies that short-range
mutations are sufficiently strong  to cancel out the dependence on the
genetic vicinity of strains with the same phenotype
to strains with different fitness. This does not
happen for the very rough landscape case, 
even though the linear relation is satisfied
in average. 

In order to obtain a  quantitatively correct prediction, one has to
consider that the resulting slope $\mu_{e}$ is related to the second
largest eigenvalue $\lambda_1$ of the mutation matrix by $\mu=1-\lambda_1$. 
When the phenotype only
depends on the ``magnetic'' term $\mathcal{H}$, in the asymptotic
state the short-range mutations connects group of equal fitness. Thus
$\boldsymbol{A}$ and $\boldsymbol{M}_s$ commutes and indeed,
 in the limit $\mu_s\rightarrow 0$,
$\mu_{e}$ tends towards the expected values $2\mu_s/L$ of 
Eq.~(\ref{Msspectrum}).

In the opposite case, when the phenotype depends on the disordered
term $\mathcal{K}$, the application of the matrix $\boldsymbol{A}$ ``rotates''
the distribution $\boldsymbol{p}$ in a way which is practically random with
respect to the Fourier eigenvectors of $\boldsymbol{M}_s$. Thus, the effective
second eigenvalue of the mutation matrix is given by $1-\mu_s$, 
obtained averaging over all the eigenvalues of Eq~(\ref{Msspectrum}). 
Consequently, we obtain  $\mu_{e} \simeq \mu_s$, 
but clearly this convergence is
quite slow, and is observed only in the limit $\mu_s \gg c \rightarrow
0$. 

When only the $\mathcal{J}$ is present, one observes an intermediate
case, which converges very slowly to the disordered case for
$L\rightarrow \infty$. 

The simultaneous application of all three terms or the mixing of long
and short-range mutations has an addictive
effect, at least at first order. 

\subsection{A toy model for the error threshold and the mutation meltdown}
\label{sec:Toy}

Before going in deep studying a general model of an evolving ecosystem
that includes the effect of competition (co-evolution), let us discuss
a simple model~\cite{Toy98} that presents two possible mechanisms of 
escaping from a
local optimum, i.e.  the error threshold and the Muller's ratchet.

We consider a \emph{sharp peak landscape}: the 
phenotype $u_0=0$, corresponding to the master sequence 
genotype $x=0\equiv(0,0,\dots)$ 
has higher fitness $A_0=A(0)$, and all 
other genotypes have the same, lower, fitness $A_*$.  
Due to the form of the fitness function, the dynamics of the
population is fundamentally determined by the fittest strains.
The effect of global competition for shared resource is considered introducing
a standard Verhulst factor, similar to Eq.~(\ref{logistic}).

Let us indicate
with $n_0=n(0)$ the number of individuals sharing the master sequence, 
 with $n_1=n(1)$ the
number of individuals with phenotype $u=1$ (only one bad gene, i.e. 
a binary string with all zero, except a single $1$), and with
$n_*$ all
other individuals. 
We assume also non-overlapping generations, 

During reproduction, individuals with phenotype $u_0$ can mutate, 
contributing to $n_1$, and those with phenotype $u_1$ can mutate, 
increasing  $n_*$. 
We disregard the possibility of back mutations from
$u_*$ to $u_1$ and from $u_1$ to $u_0$. 
This last assumption is
equivalent to the limit $L\rightarrow \infty$, which is the case for existing
organisms. We consider only short-range mutation with probability
$\mu_s$.
Due to the assumption of large $L$, the multiplicity factor of
mutations from $u_1$ to $u_*$ (i.e. $L-1$)
is almost the same of that from $u_0 $
to $u_1$ (i.e. $L$).  

We  assume a finite (and constant) 
carrying capacity $K$ of the environment, 
assuming that the effective reproduction 
rate of a population is proportional to the Verhulst factor
$1-N/K$, where $N=n_0+n_1+n_*$ is the total number of individuals.
 
The evolution equation of the population is
\begin{equation}
\begin{array}{rcl}
 n'_0 &=&\left(1-\dfrac{N}{K}\right)(1-\mu_s) A_0 n_0, \\
 n'_1 &=&\left(1-\dfrac{N}{K}\right)\bigl(
  (1-\mu_s) A_* n_1+\mu_s A_0 n_0\bigr),\\
 n'_* &=&\left(1-\dfrac{N}{K}\right)  A_* (n_*+\mu_s n_1).
\end{array}
\label{toy:n}
\end{equation}

The evolution equation for the total population is 
\[
	N'=\overline{A}N\left(1-\dfrac{N}{K}\right),
\]
where 
\[
	\overline{A}=\dfrac{A_0 n_0 + A_* (n_1+n_*)}{N}
\]
is the average fitness of the population.

The steady state of Eq.~(\ref{toy:n}) 
is given by $\boldsymbol{n}'=\boldsymbol{n}$. 
There are three possible fixed points:
\begin{equation}
	P_1 = \left\{\begin{array}{rcl}
		n_0^{(1)} &=&0, \\
		n_1^{(1)} &=&0,\\
		n_*^{(1)} &=&0,\\
		N^{(1)}&=&0;
	\end{array}\right.
	\label{P1} 
\end{equation}
\begin{equation}
	P_2 = \left\{\begin{array}{rcl}
		n_0^{(2)} &=&0, \\
		n_1^{(2)} &=&0,\\
		n_*^{(2)} &=&N^{(2)},\\
		N^{(2)} &=& K\left(1-\dfrac{1}{\overline{A_*}}\right);
	\end{array}\right. 
	\label{P2}
\end{equation}
and
\begin{equation}
	P_3 = \left\{\begin{array}{rcl}
		n_0^{(3)} &=&N^{(3)}\dfrac{(1-\mu_s) A_0 -A_*}{A_0 -A_*}, \\
		n_1^{(3)} &=&N^{(3)}\dfrac{\mu_s}{1-\mu_s}\dfrac{ A_0(q A_0
			-A_*)}{(A_0 -A_*)^2}, \\
		n_*^{(3)} &=&N^{(3)}\dfrac{\mu_s^2}{1-\mu_s} \dfrac{A_0A_*}{(A_0
			-A_*)^2},\\
		N^{(3)} &=& 1-\dfrac{1}{A_0(1-\mu_s)}.
	\end{array}\right.
	\label{P3} 
\end{equation}

The fixed point $P_1$ corresponds to extinction of the whole
population, i.e.\ to
mutational meltdown (MM). It is
trivially stable if $A_0<1$, but it can acquire stability also if $A_0>1$, $A_*
<1$  and 
\begin{equation}
	\mu_s > 1-\dfrac{1}{A_0}.
	\label{MR}
\end{equation} 

The fixed point $P_2$ corresponds to a distribution in which the master sequence
has disappeared even if it has larger fitness than other phenotypes.
This effect is usually called Muller's ratchet (MR).
 The point $P_2$ is stable for $A_0>1$,  $A_*
>1$ and 
\begin{equation}
	\mu_s > \dfrac{A_0/A_*-1}{A_0/A_*}.
	\label{MM}
\end{equation}

The fixed point $P_3$ corresponds to a coexistence of all phenotypes. 
It is stable in the rest of cases, with $A_0>1$. 
The asymptotic distribution, however, can assume two very different shapes.
In the 
quasi-species (QS) distribution, 
the master sequence phenotype is more abundant
than other phenotypes; after increasing the mutation rate, however, the numeric
predominance of the master sequence is lost, an effect that can be denoted error
threshold (ET).
The transition between these two regimes is given by $n_0=n_1$, i.e.\ 
\begin{equation}
	\mu_s = \dfrac{A_0/A_*-1}{2A_0/A_*-1}. 
	\label{ET}
\end{equation}
Our definition of the error threshold transition
needs some remarks: in Eigen's 
original work~\cite{Eigen71,Eigen:quasispecies} the error
threshold is
located at the maximum mean Hamming distance, which
corresponds to the maximum spread of population. In the limit of very
large genomes these two definitions agree, since the transition
becomes very sharp~\cite{Galluccio}. See also
Refs.~\cite{Baake1,Baake3}.

\begin{figure}[t]
\centerline{
\psfig{figure=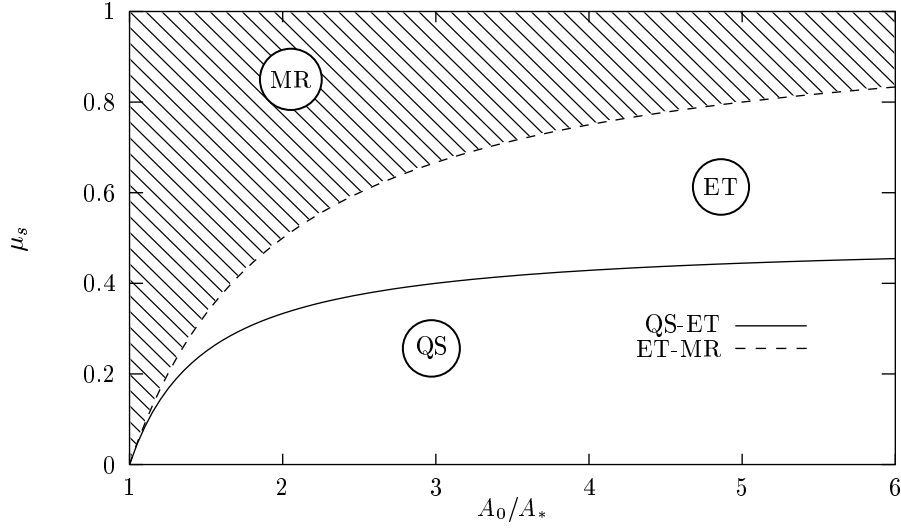,width=12cm}
}
\caption{Phase diagram for the error threshold and Muller's ratchet
transitions ($A_*>1$).
 MR refers to the Muller's ratchet phase
ET  to the error threshold distribution and QS to
 quasi-species distribution. The phase boundary between the 
 Muller's ratchet effect and
 the error threshold distribution (Eq.~(\protect{\ref{MR}})) is marked ET-MR;
the phase boundary between 
 the error threshold and the quasi-species 
 distribution (Eq.~(\protect{\ref{ET}})) is marked QS-ET. 
 }
\label{fig:toy1}
\end{figure}

In Figure~\ref{fig:toy1} we reported the phase diagram of model (\ref{toy:n})
for $A_*>1$ (the population always survives).
 There are three regions:  for a 
low mutation probability $\mu_s$ and high selective advantage $A_0/A_*$ of the
master sequence, the distribution has the quasi-species form (QS); 
increasing $\mu_s$ the distribution undergoes the error threshold (ET) effect;
finally, for very high mutation probabilities, the master sequence disappears
and we enter the Muller's ratchet (MR) region~\cite{Malarz,Baake}.

\begin{figure}
\centerline{
\psfig{figure=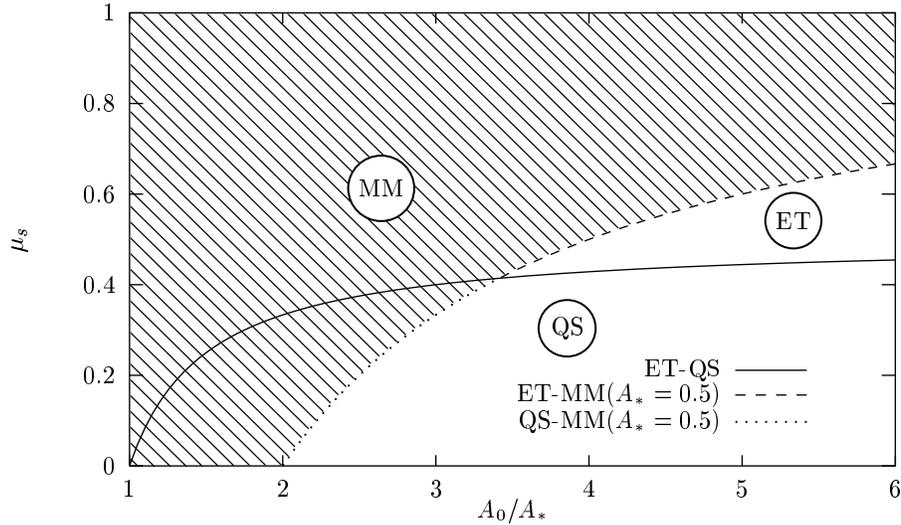,width=12cm}
}
\caption{Phase diagram for the mutational meltdown extinction,
 the error threshold and the
quasi-species distributions ($A_*<1$).
 MM refers to the mutational meltdown phase,
ET  to the error threshold distribution and QS to
 quasi-species distribution. The phase boundary between the 
 Mutational meltdown effect and
 the error threshold distribution (Eqs.~(\protect{\ref{MM}})
 and (\protect{\ref{ET-MM}})) 
 is marked ET-MM;
the phase boundary between 
 the mutational meltdown and the quasi-species 
 distribution (Eqs.~(\protect{\ref{MM}}) and (\protect{\ref{QS-MM}})) 
 is marked QS-MM. 
 }
\label{fig:toy2}
\end{figure}

In Figure~\ref{fig:toy2} we illustrate the phase diagram 
 in the case $A_*=0.5$. For a 
low mutation probability $\mu_s$ and high selective advantage $A_0/A_*$ of the
master sequence, again one observes a quasi-species distribution (QS), while 
for sufficiently large $\mu_s$ there is the extinction of the whole population 
due to the mutational meltdown (MM) effect.  
The transition between the QS and MM phases can occur directly, for 
\begin{equation}
	A_0/A_*< \dfrac{1-\sqrt{1-A_*}}{A_*}
	\label{QS-MM}
\end{equation}
 (dotted QS-MM line in Figure): 
during the transient before
extinction the distribution keeps the QS form. For  
\begin{equation}
	A_0/A_* > \dfrac{1-\sqrt{1-A_*}}{A_*}
	\label{ET-MM}
\end{equation}
one has first the error threshold transition
(QS-ET line in Figure),
and then one observes extinction due to the mutational meltdown effect (dashed
ET-MM line in Figure). This mutation-induced
extinction has been investigated numerically in Ref.~\cite{Malarz}. 

\begin{figure}
\centerline{
\psfig{figure=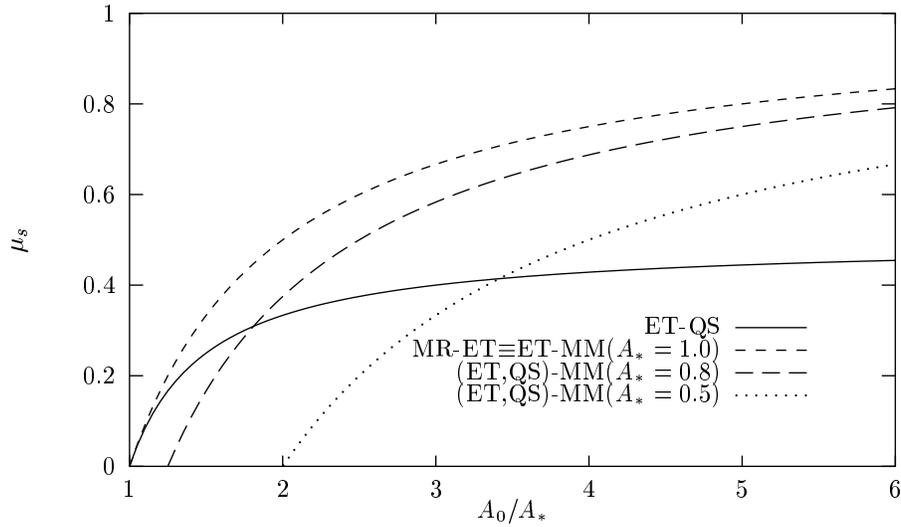,width=12cm}
}
\caption{Phase diagram for the error threshold and mutational meltdown
transitions for some values of $A_*$. 
ET-QS refers to the Error threshold transition,
 Eq.~(\protect\ref{ET}),
QS-MM to the mutational meltdown extinction without the error threshold
transition, 
 Eqs.~(\protect\ref{MM}) and (\protect{\ref{QS-MM}}),
ET-MM to the mutational meltdown extinction after the error threshold
transition, Eqs.~(\protect\ref{MM}) and  (\protect{\ref{ET-MM}}).
The line MR-ET marks the Muller's ratchet  boundary,  Eq.~(\protect\ref{MR}),
which  coincides with   
the mutational meltdown (MM) boundary for $A_*=1$. 
 }
\label{fig:toy3}
\end{figure}

We finally report in Figure~\ref{fig:toy3} the phase diagram of the model in the
$A_*<1$ case, for some values of $A_*$. Notice that for $A_*=1$ the mutational
meltdown effect coincides with the Muller's ratchet one. 

Here the mutation probability $\mu_s$ is defined on a per-genome basis. If one
considers a fixed mutation probability $\mu$ per genome element, one 
has $\mu_s \simeq L \mu$, where $L$ is the genome length. 
Thus, it is possible to trigger 
these phase transitions by increasing the genome length.

\subsection{Evolution in a phenotypic space}
\label{sec:PhenotypicEvolution}

In the previous section we have obtained some information about the
stability of a quasi-species distribution. In the following we want to
study the stability of a distribution formed by more than one
quasi-species, i.e.\ the speciation phenomenon. Before doing that we
need to know the shape of a quasi-species given a static fitness
landscape. Some analytical results can be obtained by 
considering the dynamics
only in the phenotypic space. 

We assume that the phenotypic index $u$ ranges between $-\infty$ and
$\infty$
in unit steps (the fitness landscape provides that only a finite range
of the phenotypic space is viable), and  that  
mutations connect 
phenotypes at unit distance; 
the probability of observing a mutation per unit of time is $\mu$. 
The mutational matrix $M(u,v)$ has the form:
\[
	 M(u,v) = 
	 \left\{\begin{array}{ll}
 	 	\mu & \mbox{if $|u,v|=1$,} \\
 		1-2\mu & \mbox{if $u=v$,}\\
		0&\mbox{otherwise.}
	\end{array}\right.
\]

Let us  consider as before the evolution of phenotypic distribution $p(u)$,
that gives the probability of observing the phenotype $u$. 
As before the whole distribution is denoted by $\boldsymbol{p}$.

Since we are interested in studying the speciation transition in this section we
 consider model {\bf (b)} (Section~\ref{const}), the results of this 
analysis can be partly
applied to model {\bf (a)}  as discussed at the end of this section.

Considering a phenotypic linear space and the non-overlapping generations 
limit ($\pi \rightarrow 0$) we get from Eq.(\ref{mutsel})

\[
	p'(u)=
	\dfrac{(1- 2 \mu)A(u,\boldsymbol{p})p(u) + \mu (A(u+1,\boldsymbol{p})p(u+1)+
	A(u-1,\boldsymbol{p})p(u-1)}{\overline A}.   
\]

In the limit of continuous phenotypic space, $u$ becomes a real number and 
\begin{equation}
p'(u)= \dfrac{1}{\overline{A}}\left(
	A\bigl(u,\boldsymbol{p}\bigl)p(u) 
	+\mu \dfrac{\partial^2A\bigl(u,\boldsymbol{p}\bigl)p(u)}{\partial
	u^2}
 \right), \label{evol1}
\end{equation}

with
\begin{equation}
\int_{-\infty}^{\infty} p(u)du =1, \qquad \int_{-\infty}^{\infty} A(u,\boldsymbol{p})p(u) du=\overline{A}.
\label{norm}
\end{equation}

Eq.~(\ref{evol1}) has the typical form of 
 a nonlinear diffusion-reaction equation.
The numerical solution of this equation shows that a stable asymptotic 
distribution exists for almost all initial conditions.

The fitness $A(u,\boldsymbol{p})=\exp (H(u,\boldsymbol{p}))$ can be written 
as before, with
\[
	H(u,\boldsymbol{p})=V(u)+\int_{-\infty}^{\infty} J(u,v) p(v) dv .\label{fitness}
\]

Before studying the effect of competition and the speciation transition
let us derive the exact form of
$p(u)$ in case of a smooth and sharp static fitness landscape.

\subsubsection{Evolution near a smooth and sharp maximum} 
\label{sec:smooth}

In the presence of a single maximum  
the asymptotic distribution is given by one quasi-species
centers around the global maximum of the static landscape. 
The effect of a finite mutation rate is simply that of broadening
the distribution from a delta peak to a bell-shaped curve.

We are interested in deriving the exact form of the 
asymptotic distribution near the maximum. 
We take a static fitness $A(u)$
with a smooth, isolated maximum for $u=0$ ({\it smooth maximum} approximation).
Let us assume that 
\begin{equation}
	A(u)\simeq A_{0}(1-au^{2}), 
	\label{pot}
\end{equation}
where $A_0 = A(0)$.
Substituting $\exp (w)=Ap$ in Eq.~(\ref{evol1}) we have (neglecting to indicate
the phenotype  $u$, and using primes to denote differentiation with respect
to it): 
\[
	\dfrac{\overline{A} }{A}=1+\mu ({w'}^{2}+w''), 
\]
 and approximating $A^{-1}=A_0^{-1}\left( 1+au^{2}\right) $, we have 
\begin{equation}
\dfrac{\overline{A} }{A_{0}}(1+au^{2})=1+\mu ({w'}^{2}+w''). \label{alphaA0}
\end{equation}
A possible solution is 
\[
	w(u)=-\dfrac{u^{2}}{2\sigma ^{2}}. 
\]
Substituting into Eq. (\ref{alphaA0}) we finally get 
\begin{equation}
	\dfrac{\overline{A} }{A_{0}}=\dfrac{2+a\mu - 
	\sqrt{4a\mu +a^{2}\mu ^{2}}}{2}. 		
	\label{smooth:app}
\end{equation}
Since $\overline{A} /A_{0}$ is less than one we have
 chosen the minus sign. In the limit $a\mu \rightarrow 0$ (small mutation rate
and smooth maximum), we have 
\begin{equation}
	\dfrac{\overline{A} }{A_{0}}\simeq 1-\sqrt{a\mu } \label{aveA}
\end{equation}
and
\begin{equation}
	\sigma ^{2}\simeq \sqrt{\dfrac{\mu }{a}}. \label{sigma}
\end{equation}

The asymptotic solution is 
\begin{equation}
	p(u)= \dfrac{1+au^{2}}{\sqrt{2\pi }\sigma (1+a\sigma ^{2})}\exp \left(
	-\dfrac{u^{2}}{2\sigma ^{2}}\right), \label{phenotypicp}
\end{equation}
so that $\int p(u)du=1$. The solution is then a bell-shaped curve, its
width $\sigma$ being determined by the combined effects
of the curvature $a$ of maximum and the mutation rate $\mu$.

For completeness, we study here also the case of a {\it sharp maximum},  
for which  $A(u)$ varies considerably with $u$. In this case 
the growth rate of less fit strains has a 
large contribution from the mutations of fittest strains, 
while the reverse flow is negligible, thus
\[
	p(u-1)A(u-1) \gg p(u)A(u) \gg p(u+1)A(u+1) 
\]
neglecting last term, and substituting  $q(u)=A(u)p(u)$ in Eq.~(\ref{evol1})
we get: 
\begin{equation}
	\dfrac{\overline{A}}{A_0}  = 1-2\mu \qquad \mbox{for $u=0$}\label{map0}
\end{equation}
and 
\begin{equation}
 	q(u) =\dfrac{\mu}{\left(\overline{A} A(u) 
 	-1+2\mu\right)} q(u-1) \qquad \mbox{for
 	$u>0$} \label{map}
\end{equation}

Near $u=0$, combining Eq.~(\ref{map0}), Eq.~(\ref{map})and Eq.~(\ref{pot})), 
we have
\[
	q(u) =\dfrac{\mu}{(1-2\mu)a u^{2}} q(u-1). 
\]

In this approximation the solution is
\[
	q(u) = \left(\dfrac{\mu}{1-2\mu a}\right)^u \dfrac{1}{(u!)^2},
\]
and 
\[
	y(u) = A(u)q(u) \simeq \dfrac{1}{A_0}(1+a u^2)
	\left(\dfrac{\mu A_0}{\overline{A}
	a}\right)^u \dfrac{1}{u!^2}. 
\]

\begin{figure}
\psfig{figure=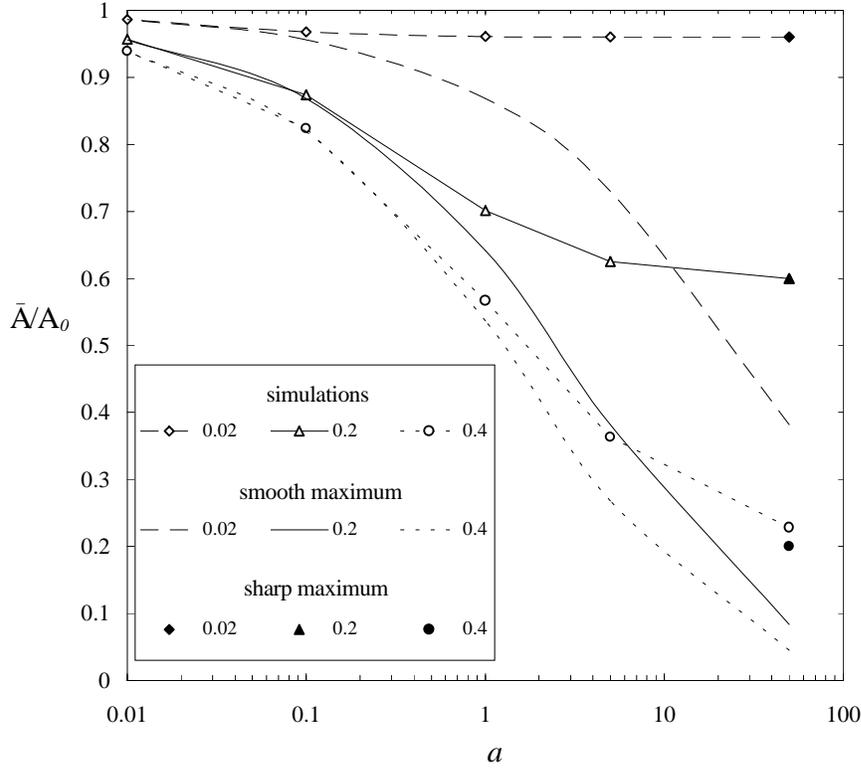,width=12cm}
\caption{Average fitness $\overline{A}/A_{0}$ versus the coefficient $a$,
of the fitness function, Eq.~(\ref{pot}),  for some values of the mutation
rate $\mu$. Legend: {\it numerical solution} corresponds to the numerical solution of 
Eq.~(\ref{evol1}),  
{\it smooth maximum} refers to  Eq.~(\ref{smooth:app}) and {\it sharp maximum}
to Eq.~(\ref{map0})}
\label{alpha}
\end{figure}

We have checked the validity of these approximations by solving numerically   
Eq.~(\ref{evol1}); 
the comparisons are shown 
in Figure~(\ref{alpha}).
We observe that the {\it smooth maximum} approximation agrees with the 
numerics for small values of $a$, 
 when $A(u)$ varies slowly with $u$, while the {\it
sharp maximum} approximation  agrees with the  numerical results for large
 values of  $a$, when small variations of $u$ correspond to large
variations of $A(u)$.

\subsection{Coexistence on a static fitness landscape}
\label{sec:Coexistence}

We study here the conditions for which more than one
quasi-species can coexist  on a static
fitness landscape without competition.

Let us assume that the fitness landscape has several distinct peaks,
and that any peak can be approximated by a quadratic function near its
maximum. 
For
small but finite mutation rates, as shown by Eq.~(\ref{phenotypicp}),
the distribution around an isolated maxima is a bell curve, whose
width is given by Eq.~(\ref{sigma}) and average fitness by
Eq.~(\ref{aveA}). Let us call thus distribution a quasi-species, and
the peak a niche.

If the niches are separated by a distance greater
than $\sigma$, a superposition of quasi-species~(\ref{phenotypicp}) 
is a solution of 
Eq.~(\ref{mutsel}). Let us number the quasi-species with the
index $k$:
\[
	p(u) = \sum_k p_k(u);
\]
each $p_k(u)$ is centered around $u_k$ and has average fitness
$\overline{A}_k$.  The condition for the coexistence of two
quasi-species $h$
and $k$ is
$\overline{A}_h = \overline{A}_k$ (this condition can be extended to
any number of quasi-species). 
In other terms one can
say that in a stable environment the fitness of all individuals is the same,
independently on the species. 
 This is a sort of global competition among all
quasi-species, due to the conservation of probability (finite carrying
capacity).

\begin{figure}
\centerline{\psfig{figure=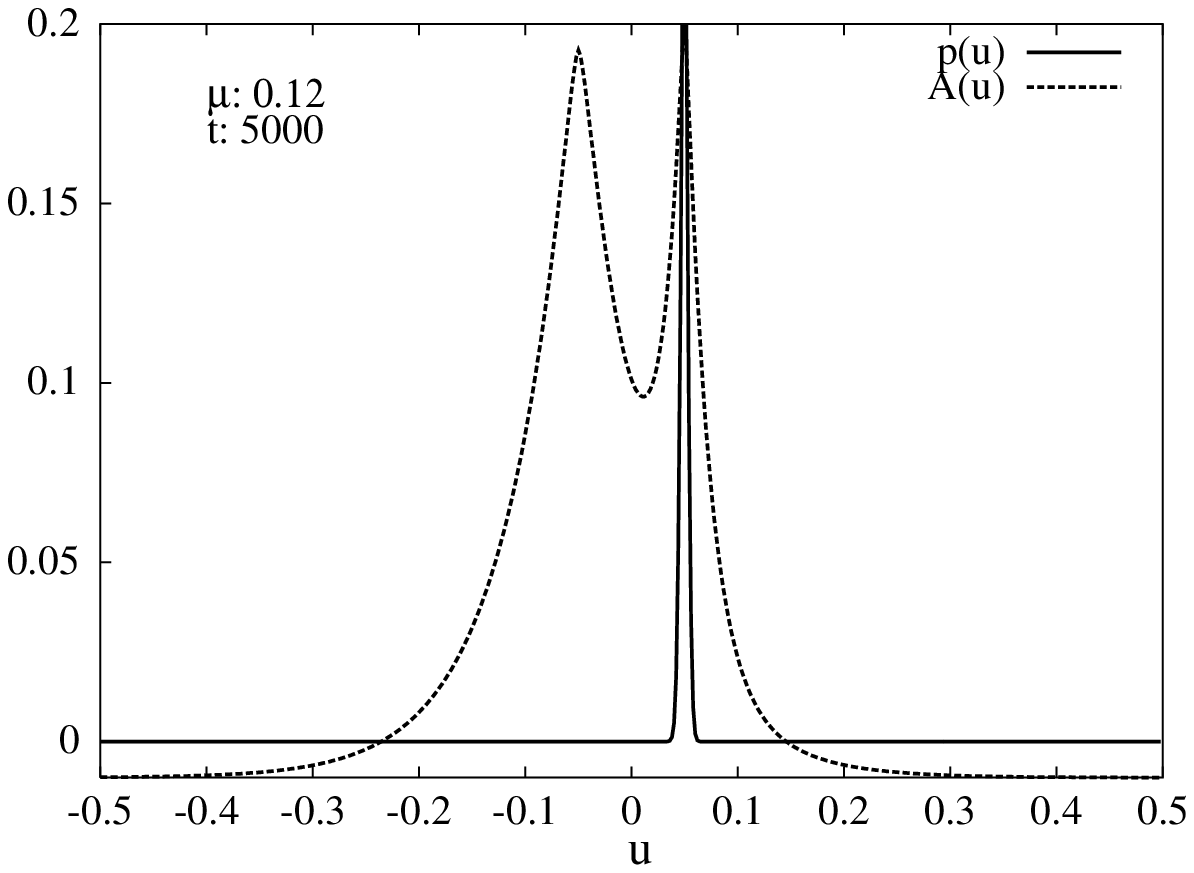,width=8cm}
\psfig{figure=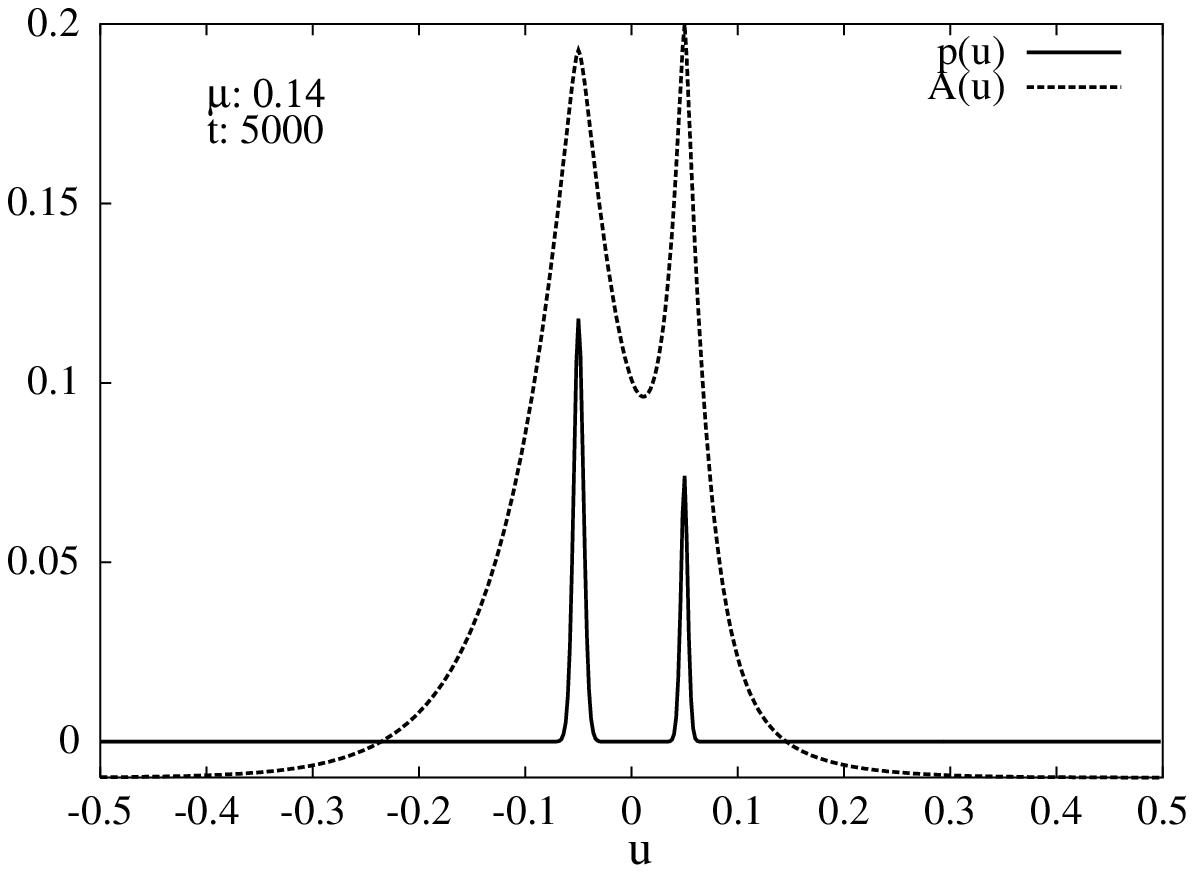,width=8cm}}
\caption{\label{figure:mu-dep} Mutation induced speciation.
A two peaks static fitness landscape, increasing the mutation rate 
we pass from a single quasi-species population (left, $\mu=0.12$) to the coexistence of
two quasi-species (right, $\mu=0.14$).  
}
\end{figure}

\begin{figure}
\centerline{\psfig{figure=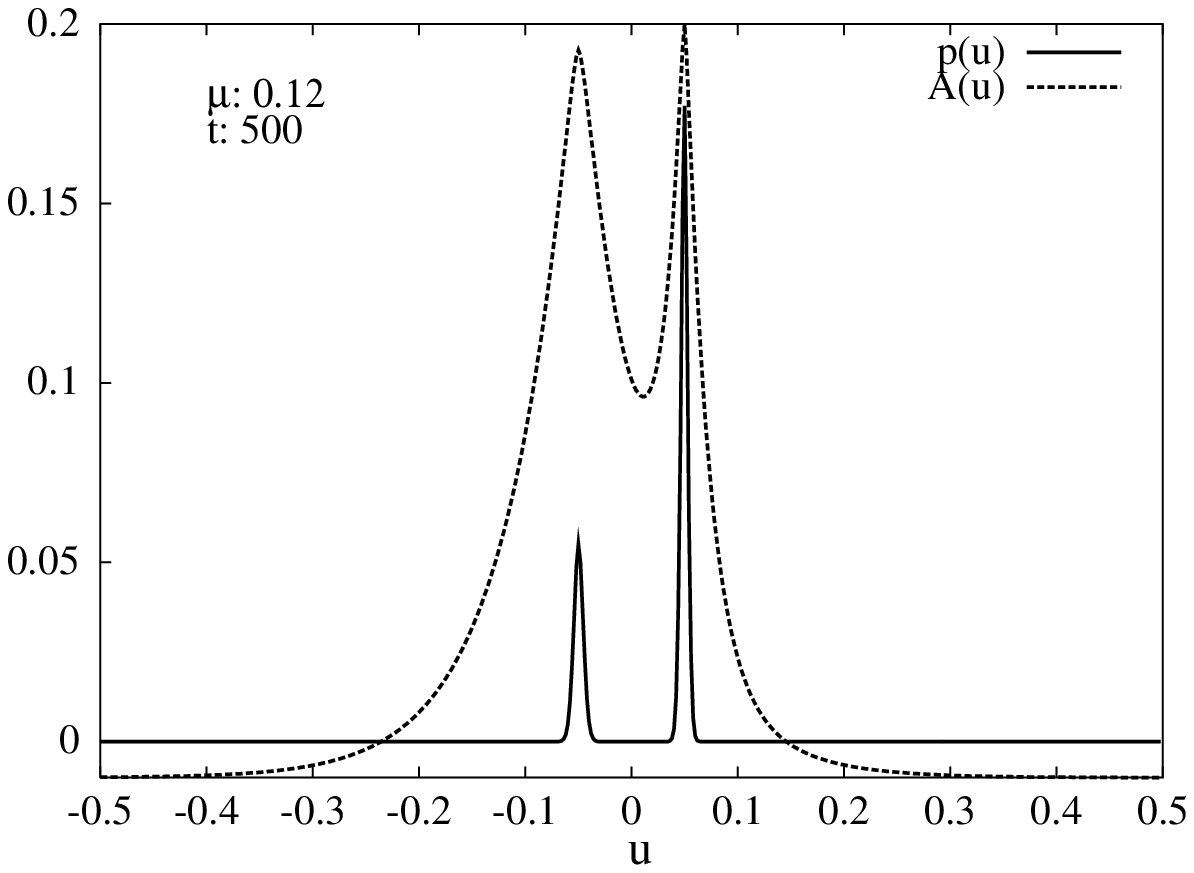,width=8cm}
\psfig{figure=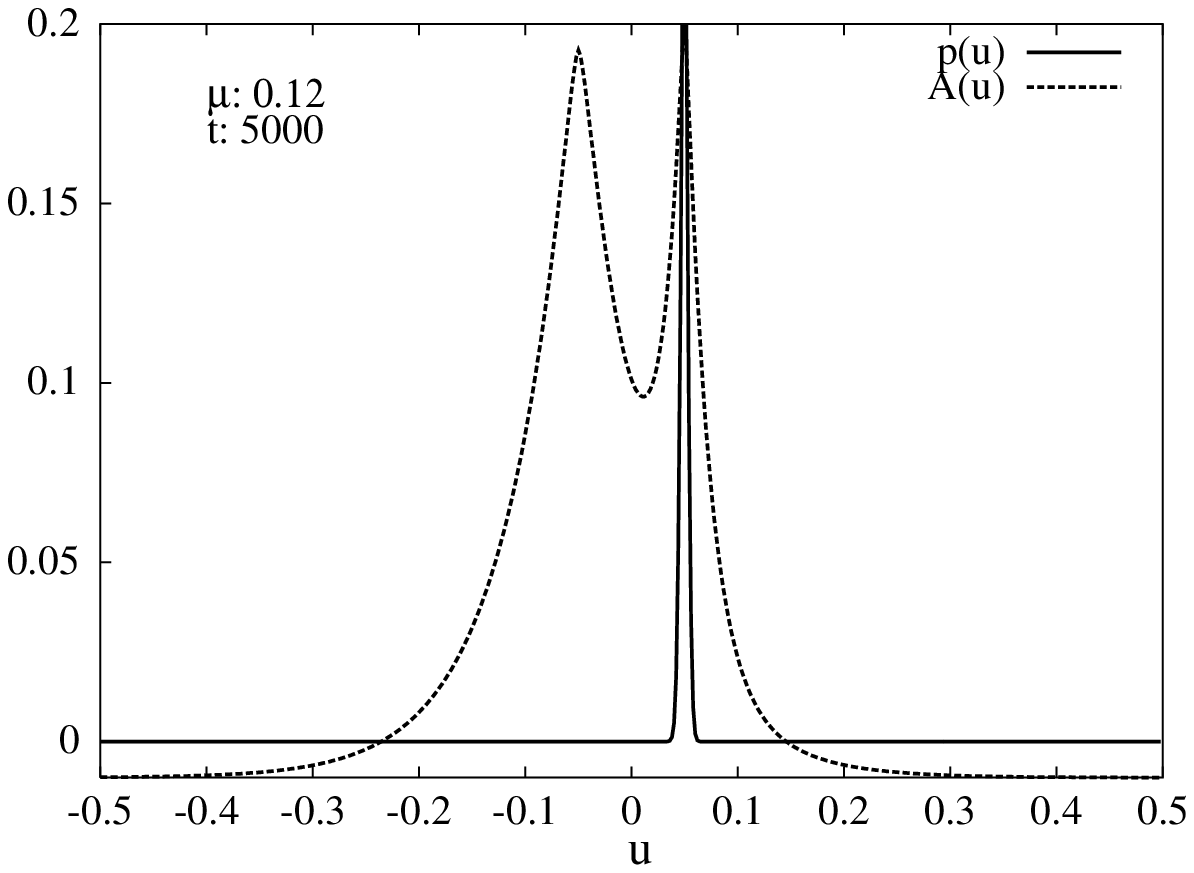,width=8cm}}
\caption{\label{figure:transient} Evolution in a two peaks 
static fitness landscape, after $500$ (left) and $5000$ (right) time steps.
For a transient period of time the two species co-exist, 
but in the asymptotic limit
only the fittest one survives.
}
\end{figure}

Since the average fitness~(\ref{aveA}) of a
quasi-species depends
on the height $A_0$ and the curvature $a$ of the niche, one can have
coexistence of a sharper niche with larger fitness 
together with a broader niche with lower fitness, as
shown in Figure~\ref{figure:mu-dep}. However, this
coexistence depends crucially on the mutation rate $\mu$. If $\mu$ is
too small, the quasi-species occupying the broader niche disappears; if
the mutation rate is too high the reverse happens. In this case, the
difference of fitness establishes the time scale, which can be quite
long. In presence of a fluctuating environment, these time scales can
be long enough that the extinction due to global competition is not
relevant. A transient coexistence is illustrated in
Figure~\ref{figure:transient}.
One can design a
special form of the landscape that allows the coexistence for a finite
interval of values of $\mu$, but certainly this is not a generic case.

We  show in the following 
that the existence of a degenerate effective fitness 
is a generic case in the presence of competition. 

\section{Evolution on a dynamic landscape: the role of competition}
\label{sec:DynamicalEvolution}

\subsection{Speciation in the phenotypic space}
\label{sec:PhenotypicSpeciation}

In this section we introduce a new factor in our model ecosystem: a
short-range (in phenotypic space) competition among individuals. As
usual, we start the study of its consequences by considering the
evolution in phenotypic space~\cite{Bagnoli:prl,Bagnoli:ecal}. 

We assume that the static fitness $V(u)$, when not flat, is a 
linear decreasing function of the phenotype $u$
except in the vicinity of $u=0$, where it has a quadratic maximum:
\begin{equation}
	V(u) = V_0 + b\left(
		1-\dfrac{u}{r} - \dfrac{1}{1+u/r}
	\right)\label{V}
\end{equation}
so that close to $u=0$ one has 
$V(u) \simeq V_0 -b u^2/r^2$ and for $u\rightarrow \infty$,
$V(u) \simeq V_0 + b(1-u/r)$. 
The master sequence is located at $u=0$.

We have checked numerically that the results are
qualitatively independent on the exact form of the static fitness, providing
that it is a smooth decreasing function. We have introduced this particular form
because it is suitable for analytical computation, but a more classic
Gaussian form can be used.

For the interaction matrix $W$ we have chosen the following  kernel $K$ 
\[
	K\left(\dfrac{u-v}{R}\right) = \exp\left(-\dfrac{1}{\alpha}
	\left|\dfrac{u-v}{R}\right|^\alpha\right).
\]
The parameter $J$  and $\alpha$ control
the intensity and the steepness of the intra-species competition,
respectively. We use a Gaussian ($\alpha=2$) kernel, for the
motivations illustrated in Section~\ref{sec:PhenotypicSpeciation}.

\begin{figure}
\centerline{\psfig{figure=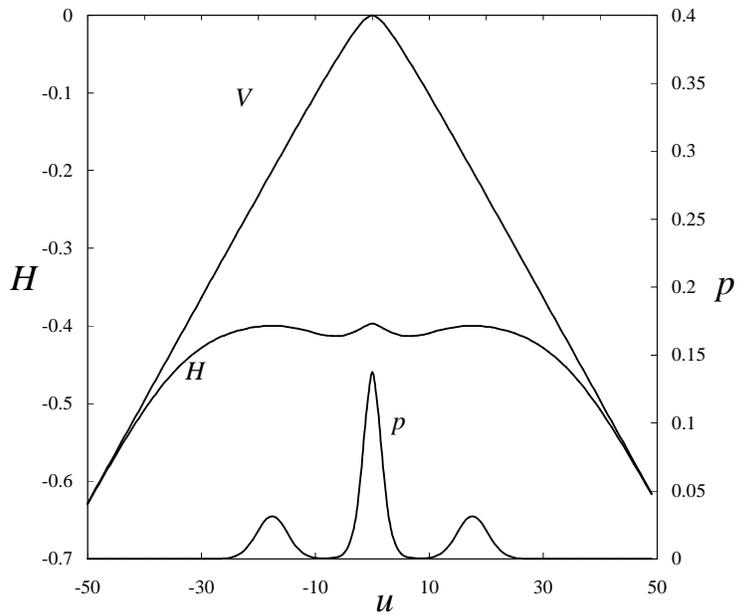,width=10cm}}
\caption{\label{Potential}
Static fitness $V$, effective fitness 
$H$,  and asymptotic distribution $p$ 
numerically computed for the following values of
parameters: $\alpha=2$, $\mu=0.01$, $V_0=1.0$, 
$b=0.04$, $J=0.6$, $R=10$, $r=3$ and $N=100$.
}
\end{figure}

For illustration, we report 
in Figure~\ref{Potential} the numerical solution of
Eq.~(\ref{mutsel}), showing
a possible evolutionary
scenario that leads to the coexistence of three quasi-species. We have
chosen the
smooth static fitness $V(u)$ of Eq.~(\ref{V})
and a Gaussian ($\alpha=2$) competition 
kernel. One can realize that 
the effective fitness  $H$ is almost 
degenerate (here $\mu>0$ and the competition effect extends on
the neighborhood of
the maxima), i.e.\ that the average fitness of all coexisting 
quasi-species is the same. 

We  now derive the conditions for the coexistence of multiple species. 
 We are interested in its asymptotic behavior in the limit
$\mu\rightarrow 0$, which is the case for actual organisms. 
Actually, the mutation mechanism is needed only to define
the genotypic distance and to populate all available niches.
Let us assume that the asymptotic distribution is formed by ${\mathcal L}$ 
quasi-species. Since $\mu\rightarrow 0$ they are approximated by delta
functions $p_k(u)=\gamma_k \delta_{u,u_k}$, $k=0, \dots, {\mathcal L}-1$,
centered at $u_k$.
The weight of each quasi species is $\gamma_k$, i.e.
\[
	 \int p_k(u) du = \gamma_k, \qquad\sum_{k=0}^{{\mathcal L}-1} \gamma_k = 1.
\]
The quasi-species are ordered such as  $\gamma_0 \ge
\gamma_1,  \dots, \ge \gamma_{{\mathcal L}-1}$. 

 The evolution equations for the $p_k$ are 
\[
	 p_k'(u) = \dfrac{A(u_k)}{\overline A} p_k(u),
\]
where $A(u) = \exp\left(H(u)\right)$ and
\[
	H(u) = 
		V(u) - J\sum_{j=0}^{{\mathcal L}-1} K\left(\dfrac{u - u_j}{R}\right) \gamma_j.
\]

The stability condition of the asymptotic distribution is 
$(A(u_k) - \overline A) p_k(u) = 0$, i.e. either
 $A(y_k) = \overline A = {\rm const}$
(degeneracy of maxima) or $p_k(u)=0$ (all other points). This supports
our assumption of delta functions for the $p_k$. 

The position $u_k$ and the weight $\gamma_k$ of the quasi-species
are given by $A(u_k) = \overline A = {\rm const}$ and 
$\left.{\partial A(u)}/{\partial u}\right|_{u_k} = 0$, or, in terms of the
fitness $H$, by
\[
	V(u_k) - J \sum_{j=0}^{{\mathcal L}-1} K\left(\dfrac{u_k-u_j}{R}\right)
		 \gamma_j = \rm{const}
\]
\[
	V'(u_k)  - \dfrac{J}{R}\sum_{j=0}^{{\mathcal L}-1}
	K'\left(\dfrac{u_k-u_j}{R}\right)
	 \gamma_j = 0
\]
where the prime in the last equation denotes differentiation with
respect to $u$. 

Let us compute the phase boundary for coexistence of three species for two
kinds of kernels: the exponential ($\alpha=1$)
and the Gaussian ($\alpha=2$) one. The diffusion kernel can be 
derived by a simple
reaction-diffusion model, see Ref.~\cite{Bagnoli:prl}. 

We assume that the static fitness $V(u)$ of Eq.~(\ref{V}). 
Due to the symmetries of the problem, we have the master quasi-species at
$u_0=0$ and, symmetrically,
two satellite quasi-species at $u=\pm u_1$. Neglecting the mutual influence of
the two marginal quasi-species, and considering that $V'(u_0) =
K'(u_0/R)=0$, 
$K'(u_1/R) = -K'(-u_1/R)$, $K(0)=J$ 
and that the three-species threshold is given by $\gamma_0=1$ and $\gamma_1=0$,
 we have 
\[
	\tilde{b}\left(1-\dfrac{\tilde{u}_1}{\tilde{r}}\right) 
				- K(\tilde{u}_1) = -1,  
\]
\[
	\dfrac{\tilde{b}}{\tilde{r}} + K'(\tilde{u}_1) = 0.
\]
where $\tilde{u}=u/R$, $\tilde{r} = r/R$ and $\tilde{b} = b/J$. 
We introduce the parameter $G=\tilde{r}/\tilde{b}=
(J/R)/(b/r)$, that is the ratio of two
quantities, the first one related to the strength of inter-species interactions
($J/R$) and the second to intra-species ones ($b/r$).

In the following we drop the tildes for convenience.
Thus
\[
 r - z - G \exp\left(-\dfrac{z^\alpha}{\alpha}\right) = -G,
\]
\[
 G z^{\alpha-1}\exp\left(-\dfrac{z^\alpha}{\alpha}\right) = 1,
\]

For $\alpha=1$ we have the coexistence condition
\[
 \ln(G) = r -1 + G.
\]
The only parameters that satisfy these equations are $G=1$ and $r=0$,
i.e.\ a
 flat landscape ($b=0$) with infinite range interaction ($R=\infty$). 
Since the coexistence region reduces to a single point,
it is suggested that $\alpha=1$ is a marginal case.
Thus for less steep potentials, such as power law decrease, 
the coexistence condition is supposed not to be
fulfilled.  

For $\alpha=2$ the coexistence condition is given by
\[
	z^2 -(G+r)z + 1 = 0,
\]
\[
	Gz\exp\left(-\dfrac{z^2}{2}\right) = 1.
\]
One can solve numerically this system and obtain the boundary 
$G_c(r)$ for the coexistence. In the limit $r \rightarrow 0$ (almost 
flat static fitness) one has 
\begin{equation}
	G_c(r) \simeq G_c(0) - r \label{Gc}
\end{equation}
with $G_c(0) = 2.216\dots$. 
Thus for $G>G_c(r)$ we have coexistence of three or more quasi-species, while 
for $G<G_c(r)$ only the fittest one survives.

\begin{figure}
\centerline{\psfig{figure=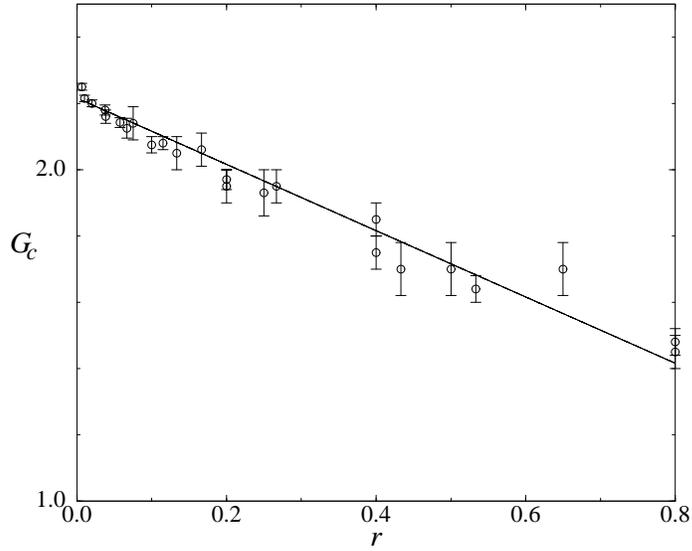,angle=270,width=10cm}}
\caption{\label{figG}
Three-species coexistence boundary $G_c$ for $\alpha=2$. The continuous 
line represents the analytical 
approximation, Eq.~(\ref{Gc}), the circles are obtained from
numerical simulations. The error bars represent the maximum error (see text for
details).
}
\end{figure}

We have solved numerically  Eq.~(\ref{evol1}) for several
different values of the parameter $G$.
We have considered a discrete phenotypic
 space, with $N$ points, and a simple Euler
algorithm. The results, presented in Figure~\ref{figG}, are not 
strongly affected by the integration step.
The error bars are due to the
discreteness of the changing parameter $G$. 
The boundary of the multi-species phase is well approximated by Eq.~(\ref{Gc});
in particular, we have checked that this 
 boundary does not depend on the mutation rate
$\mu$, at least for $\mu < 0.1$, which can be considered
a very high mutation rate for
real organisms. The most important effect of $\mu$ is the broadening of
quasi-species curves, which can eventually merge as described in
Section~\ref{sec:smooth}.

This approximate theory to derive the condition 
of coexistence of multiple quasi-species 
still holds for the hyper-cubic genotypic space.
The different structure of genotypic space does not
change the results in the  limit $\mu \rightarrow 0$.
Moreover, the threshold between one and multiple quasi-species is
defined as the value of parameters for which the satellite
quasi-species vanish. In this case the multiplicity factor of satellite
quasi-species does not influence the competition, and thus we believe
that the threshold $G_c$ of Eq.~(\ref{Gc}) still holds in the
genotypic hyper-cubic space. 

For rule {\bf (a)} with a variable
population Eq.~(\ref{peq}), the theory still works  
substituting $G$ with  
\begin{equation}
G_a= m G = m \dfrac{J/R}{b/r}, \label{Ga}
\end{equation}
 due to the presence of $m$ factor in Eq.~(\ref{Hp}) (for a detailed analysis
see Ref.~\cite{Bagnoli:ijmpc}). 

This result is in a good agreement with numerical simulations, 
as shown in the following Section.

\subsection{Speciation and mutational meltdown in the hyper-cubic genotypic space}
\label{sec:Hypercubic}

Let us now study the consequences of evolution in presence of competition 
in the more complex genotypic space. We were not able to obtain
analytical results, so we resort to numerical simulations. Some
details about the computer code we used can be found in Appendix~B. 

In the following we  refer always to rule {\bf (a)}, that allows us to
study both speciation and mutational meltdown. 
Rule {\bf (b)} has a similar behavior for speciation transition, while, of
course, it does not present any mutational meltdown transition.
 
We considered the same static fitness landscape of Eq.~(\ref{V}), and 
the simple genotype-phenotype mapping $\mathcal{J}=\mathcal{K}=0$ in
Eq.~(\ref{phenotype}) (non-epistatic interactions among genes).   
 
\begin{figure}
\centerline{\psfig{figure=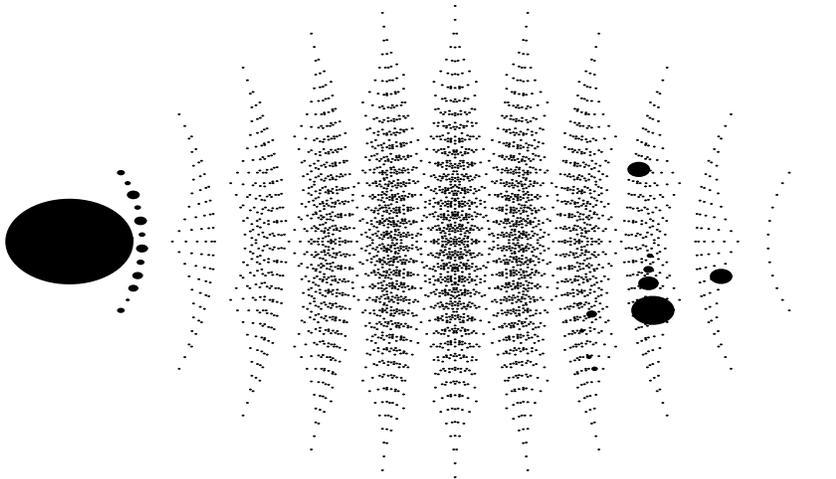,width=12cm,angle=270}}
\caption{\label{fig:Spot} Easter egg representation of quasi-species in hyper-cubic space for
$L=12$. The smallest points represent placeholder of strains (whose
population is less than $0.02$), only the
larger dots correspond to effectively populated quasi-species; the
area of the
dot is proportional to the square root of population. Parameters:
$\mu=10^{-3}$, $V_0=2$, $b=10^{-2}$, $R=5$, $r=0.5$, $J=0.28$,
$N=10000$, $L=12$.}
\end{figure}

We observe, in good agreement with the analytical approximation 
Eq.~(\ref{Gc}), that
if $G_a$ (Eq.~(\ref{Ga})) is larger than the threshold $G_c$
(Eq.~(\ref{Gc})),
several quasi-species coexist, otherwise only the master sequence
quasi-species survives. 
In Figure~\ref{fig:Spot} a distribution with multiple
quasi-species is shown.

We can characterize the speciation transition 
by means of the entropy $S$ of the asymptotic phenotypic distribution 
$p(u)$ as function of
$G_a$, 
\[ 
    S = - \sum_u p(u) \ln p(u)
\]
which increases in correspondence of the appearance of multiple
quasi-species. 

\begin{figure}
\centerline{\psfig{figure=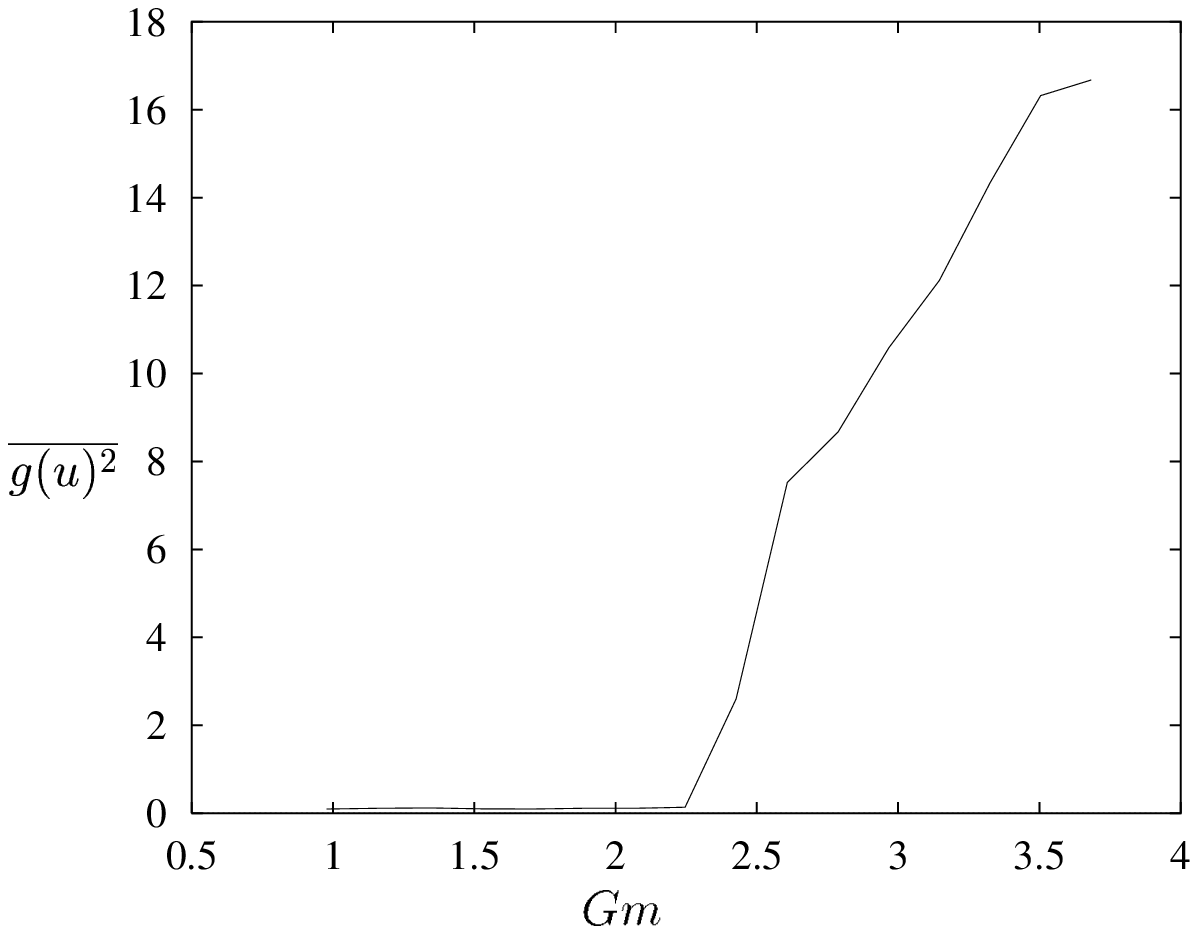,width=6.5cm}
\psfig{figure=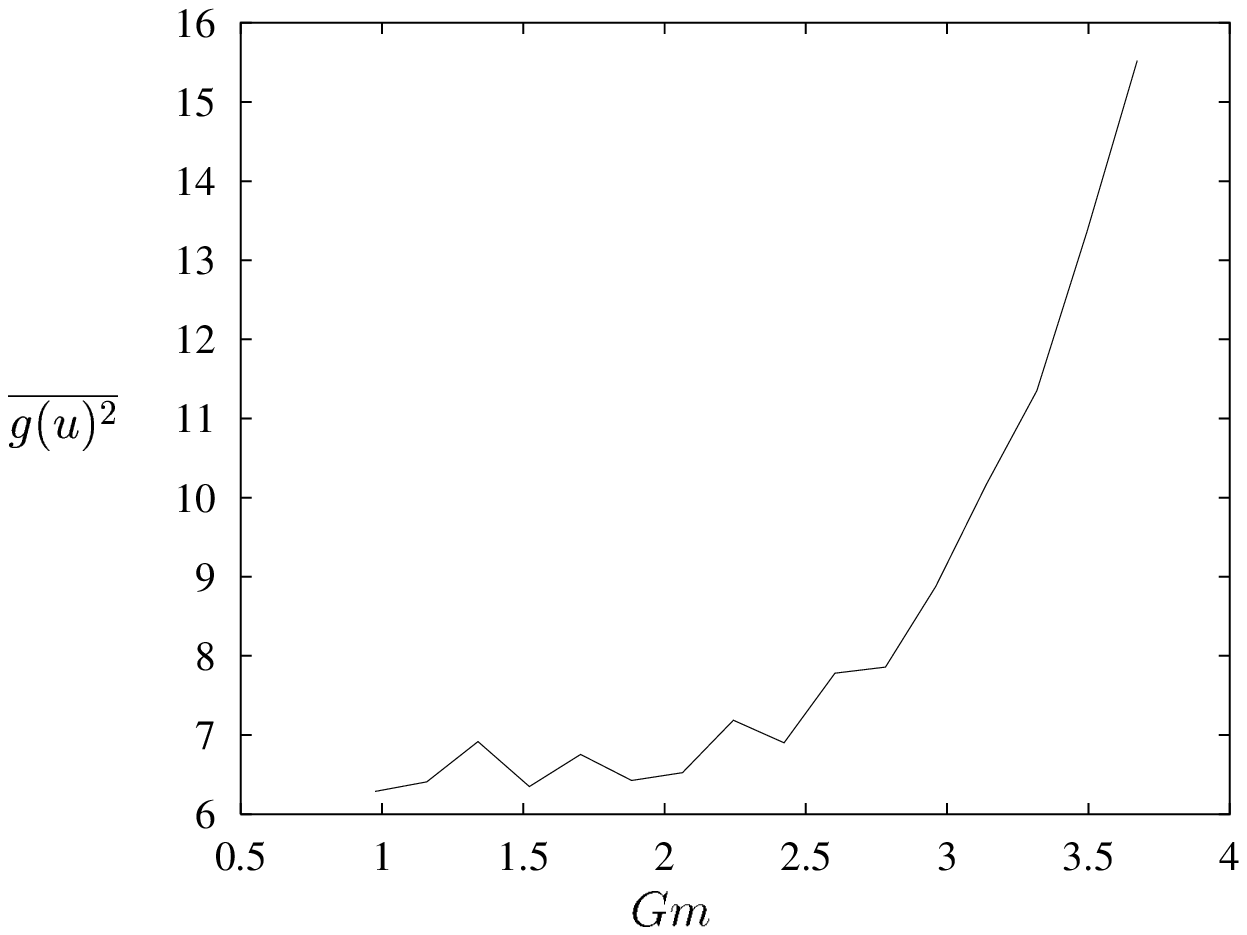,width=6.5cm}}
\caption{\label{fig:x2} Independence of 
the speciation transition by the mutation rate. 
The transition is characterized by
the average square phenotypic distance $\overline{g(u)^2}$ 
of phenotypic distribution $p(u)$, as a function of the control
parameter $G\;m$. Each point is a single run.  
Same parameters
as in Figure~\protect\ref{fig:Spot}, varying $J$ with $\mu=10^{-3}$
(left) and $\mu=5\;10^{-2}$ (right). }
\end{figure}

We locate the transition at a value $G_a \simeq 2.25$, while 
analytical approximation predicts $G_c(0.1)\simeq 2.116$.
The entropy, however, is quite sensible to
fluctuations of the quasi-species centered at the master sequence
(which embraces the
largest part of distribution), 
and it was necessary to average over several runs in order
to obtain a clear curve; for larger values of $\mu$ it was impossible
to characterize this transition. A quantity which is much less 
sensitive of fluctuations is the
average square phenotypic distance from the master sequence
 $\overline{g(u)^2}$ 
\[
 	\overline{g(u)^2}=\sum_u g(u)^2 p(u).
\]
In Figure~\ref{fig:x2} (left)  we characterize the speciation transition by
means of $\overline{g(u)^2}$, and indeed a single run was sufficient,
for $\mu=10^{-3}$. For much higher mutation rates ($\mu=5\; 10^{-2}$) the
transition is less clear, as shown in Figure~\ref{fig:x2} (right),
but one can see that the transition point is substantially
independent of $\mu$, as predicted by the analysis of speciation in the
phenotypic space, Eq.~(\ref{Gc}).

\begin{figure}
\centerline{\psfig{figure=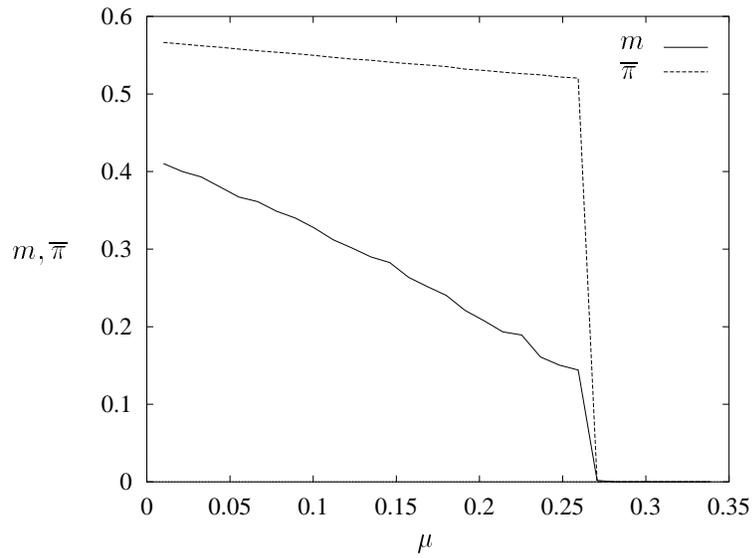,width=10cm}}
\caption{\label{fig:melt} Meltdown transition  characterized by
the total population size $m$  and the average fitness 
$\overline{\pi}$ as a function of mutation rate $\mu$. 
Here $J=0.3$,
$V_0=0.4$, $b=0.35$, $r=0.5$, $R=5$, $N=2000$, $L=8$.}
\end{figure}

Another interesting phenomenon is the meltdown transition, 
for which the spatial mean field theory predicts 
extinction  if
$\overline{\pi}\le 1/2$, Eq.~(\ref{mequation}). 
In Figure~\ref{fig:melt} we report the result of one simulation in
which the extinction  is induced by the increase of the mutation rate
$\mu$. One can notice that the transition is discontinuous, $m$ jumps
to zero from a value of about 0.15, and that the critical value of
$\overline{\pi}$ is larger that the predicted one.   
This discrepancy can be caused by fluctuations, due to the finiteness
of population.

\section{Discussion}
\label{CA:disc}
We studied some simple models of evolving asexual populations on a smooth
landscape.
After some preliminary descriptions, 
we introduced  a microscopic cellular automaton model, and we obtained its 
spatial mean-field equations. Afterwards, we studied 
the consequences of evolution in a
static fitness landscape. First of all we investigated the limit of validity of
the naive relationship between energy and selection, and temperature and
mutations, i.e.\  between the time evolution of the population and equilibrium
properties of a statistical mechanics system. We found that indeed a limit
exists for which this mapping is valid.
We then studied the 
 mutational meltdown, Eigen's 
error threshold and  Muller's ratchet phenomena introducing 
a minimal model. It is shown that the mutational meltdown may overcome the error
threshold transition, while this does not happen with the Muller's ratchet
effect. 
Finally, we derived the shape of a quasi-species and the condition of
coexistence of multiple species in a fixed fitness landscape. 
For  vanishing mutation rate, 
the coexistence of many species that share the same environment
is related to the degeneracy of maxima of the static fitness,
 that is a rather peculiar condition.
In the presence of a finite mutation rate, the highest maxima start being 
populated and more quasi-species appear. This speciation mechanism
 depends crucially on the mutation rate.
This is in contrast with the biological data, that indicate speciation 
is very little influenced by the mutation rate.

In the second part, we applied these results to the study of coexistence in
presence of competition, obtaining the conditions for robust speciation in
asexual populations.  
Our  model  includes the competition among individuals, 
and this
ingredient is  fundamental for the speciation phenomenon in
a smooth fitness landscape. 
In fact in presence of a competition term  
large ``enough'' we observe the stable coexistence of many quasi-species. 

This speciation transition does not depend on the mutation
rate provided that this rate is small. 
We are also able to
obtain  analytical approximations for the onset of both transitions. 

\section*{Acknowledgements}
We wish to thank D. Stauffer for having suggested and 
strongly urged us to write this paper. We also wish to acknowledge our
participation to the DOCS~\cite{DOCS} group in Florence.
This work was partially supported by the INFM Parallel Computing
Initiative

 
\section*{Appendix A: Properties of mutation matrices}

We discuss here some properties of the mutational matrices introduced
in Section~\ref{sec:GenotypicSpace}.
Those are circular Markov 
matrices on a hyper-cubic space.
In analogy with circular matrices in the usual space, one 
can obtain their complete
spectrum (which will be used in Section~\ref{sec:Path}) using the
analogous of Fourier transform in a Boolean hyper-cubic space. Let us
define the ``Boolean scalar product'' $\odot$:
\[
	x \odot y = \bigoplus_{i=1}^L x_i y_i,
\]
where the symbol $\oplus$ represent the sum modulus two (XOR) and the
multiplication can be substituted by an AND (which has the same effect
on Boolean quantities). This scalar product is obviously distributive
with respect to the XOR:
\[
	(x \oplus y) \odot z = (x\odot z) \oplus (y\odot z).
\]

Given a function $f(x)$ of a Boolean quantity $x \in \{0,1\}^L$, its ``Boolean 
Fourier transform'' is $\tilde f (k)$ ($k \in \{0,1\}^L$)
\[	
	\tilde f(k) = \dfrac{1}{2^L} \sum_x (-1) ^{x\odot k} f(x).
\]
The anti-transformation operation is determined by the definition of the
Kronecker delta
\[
	\delta_{k0} = \dfrac{1}{2^L} \sum_x (-1) ^{x\odot k}.
\]
One can easily verify that the Fourier vectors
\[
	\xi_k(x) = (-1)^{x \odot k} 
\]
are eigenvector of both $\boldsymbol{M}_{\ell}$ and $\boldsymbol{M}_s$, with eigenvalues
\begin{equation}	
	\begin{array}{rl}
		\lambda_0&=1,\\
		\lambda_k &= 1-\mu_{\ell}-\dfrac{\mu_{\ell}}{2^L-1},
	\end{array}
	\label{Mlspectrum}
\end{equation}
for the long-range case, and
\begin{equation}
	\begin{array}{rl}
		\lambda_0&=1,\\
		\lambda_k &= 1-\dfrac{2\mu_s d(k,0)}{L},
	\end{array}
	\label{Msspectrum}
\end{equation}
for the short-range case, where $d(x,y)$ is the Hamming distance. 
	    
\section*{Appendix B: Monte Carlo Algorithm}
We schematically describe here the code we use
to simulate the evolution of the population in the genotypic space in
presence of competition.
The source code (in C) is available upon request. 

The problem of performing a massive simulation of the ecosystem
described in this paper relies in Eq.~(\ref{fitnessH}): we have to
compute $N^2$ terms, where $N$ is the size of populations. 
Since individuals with the same phenotype contribute
equally to the competition term $W$, we can rewrite Eq.~(\ref{fitnessH})
as
\[
  H(u,\boldsymbol{n}) = V(u) + 
   \dfrac{1}{N}\sum_{v} W(u,v) n(v),
\]
where we have replaced a sum over the individuals with a
a sum over phenotypes. This solves the problem if
 the possible phenotypes are few, as for
instance if one has $\mathcal{J}=\mathcal{K}=0$ in
Eq.~(\ref{phenotype}). In this case there are $L+1$
  possible phenotypes, regardless of $N$. 
When the number of possible phenotypes is large (maybe of the order of
the population size, as happens if 
$\mathcal{K}\ne 0$), one has to avoid the summation over  
those phenotypes not present in the environment.
This may be done by keeping a
list of the actually present phenotypes, with the relative
multiplicity. A hash function
(for instance, a table) is
needed to quickly access the list. Furthermore, there is the 
 possibility of grouping the phenotypes in bins.
In practice, this is realized by addressing  
the list of phenotypes by means of a discrete phenotypic index.

\newpage

\begin{figure}
\centerline{\psfig{figure=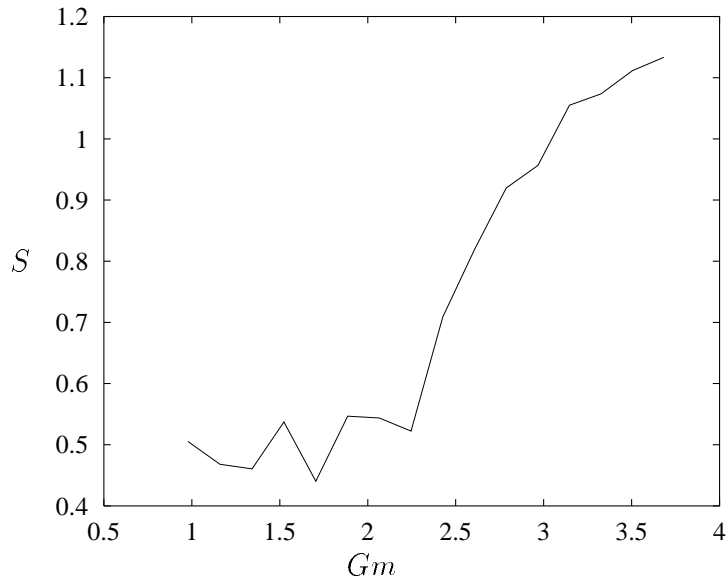,width=10cm}}
\caption{\label{fig:Entropy} The speciation transition characterized
by the entropy $S$ as a function of the control
parameter $G\;m$. Each point is an average over 15 runs.  Same parameters
as in Figure~\protect\ref{fig:Spot}, varying $J$. Errors are of the order of
fluctuations.}
\end{figure}

\end{document}